\documentstyle[aps,pra,12pt]{revtex}
\input epsf
\begin{document}
\author{Fabi\'an H. Gaioli\thanks{%
Electronic address: gaioli@iafe.uba.ar.} and Edgardo T. Garcia Alvarez}
\address{Instituto de Astronom\'\i a y F\'\i sica del Espacio, \\
C.C. 67, Suc. 28, 1428 Buenos Aires, Argentina \\
and Departamento de F\'\i sica, Facultad de Ciencias Exactas y Naturales, \\
Universidad de Buenos Aires, 1428 Buenos Aires, Argentina}
\medskip
\author{Javier Guevara}
\address{Departamento de F\'\i sica, Comisi\'on Nacional de Energ\'\i a At\'omica, \\
Avda. del Libertador 8250, 1429 Buenos Aires, Argentina \\
and Escuela de Ciencias y Tecnolog\'\i a, Universidad Nacional de San
Mart\'\i n, \\
Alem 3901, 1651 San Andr\'es, Buenos Aires, Argentina}
\title{Quantum Brownian motion }
\maketitle

\vspace{0.9in}
\begin{abstract}
We study the behavior of a subsystem (harmonic oscillator) in contact with a
thermal reservoir (finite set of uncoupled harmonic oscillators). We exactly
solve the eigenvalue problem and obtain the temporal evolution of the
dynamical variables of interest. We show how the subsystem goes to equilibrium
and give quantitative estimates of the Poincar\'e recurrence times. We
study the behavior of the subsystem mean ocuppation number in the limit of a
dense bath and compare it with the expected exponential decay law.
\end{abstract}

\newpage
\section{Introduction}
\medskip

One of the long standing paradoxical problems in theoretical physics lies in
understanding how a macroscopic system reaches equilibrium departing from
the reversible microscopical laws of nature. The simplest systems for which
the origin of irreversibility can be studied on a microscopic basis are the
linear ones \cite{Ullersma,Gordo}. In this work we investigate this behavior
for a Brownian particle in a quantum-mechanical heat bath composed of a
finite number of small oscillators. The key point of this analysis is the
consideration of the linear coupling between the Brownian particle and the
bath, which allows us to reduce the Hamiltonian to a set of
uncoupled oscillators. This model is recurrently studied in the literature
from many different approaches, such as the Langevin equation, the master
equation, and the exact solution for the evolution operator, using different
techniques. The importance of this model lies in its broad applicability in
many fields of physics: condensed matter, statistical
mechanics, quantum optics, quantum electrodynamics, quantum measurement
theory, scattering and
decay theory, etc. In the majority of previous works
statistical fluctuations, dissipation, and equilibrium tendency in linear
quantum-mechanical systems are shown to result from a projection of the
total quantum system onto a restricted subspace. A macroscopic equation is
obtained corresponding to a reduced description of the system. The
restriction to the Markovian approximation and weak-coupling limit are
usually undertaken as well as the analysis is carried out in the limit of a
dense bath. Only few works are devoted to study the behavior of the system
for a finite (discrete spectrum) bath \cite{Kozak,Gruver}. In this case it
is not possible to prove convergence to an equilibrium state in the limit $%
t\rightarrow \infty $ because of the existence of Poincar\'e recurrences,
which become extremely infrequent for large systems. This is the reason why
previous works prefer to eliminate them by passing to the limit of an
infinite heat bath. However, quantitative estimates of the Poincar\'e
recurrence times for finite systems can be made in a way compatible 
with a dissipative
behavior. In this work we reinforce this fact. We study the time evolution
of a finite system departing from the exact solution of the eigenvalue 
problem without appealing to
approximations, assuming a factorizable initial condition where the bath has
reached a unique thermal equilibrium state (passive reservoir). No use is
made of coarse graining, finite memory assumptions, randomly varying
Hamiltonians or nonlinear modifications to the Schr\"odinger evolution. We
perform our calculations for an arbitrary spectral density and temperature.

The work is organized as follows. In Sec. II we introduce the model and its
exact solution. We also show a criterion to disregard a broad
class of coupling functions. The exact solution is used in Sec. III in order
to study the time evolution of the relevant variables of our problem, i.e.
the mean occupation number (or energy) and mean position of the subsystem,
the behavior of bath variables, etc. Sec. IV is devoted to derive an exact
generalized form of the quantum Langevin equation, with time-dependent
coefficients. Numerical results and their analysis are presented in
Sec. V. We show that the mean position operator performs damping
oscillations correlated with the mean energy of the subsystem, which decays
in time towards a state of equilibrium with the bath. After reaching 
this state revivals occurs periodically in
the subsystem. 
The bath remains
almost unaltered in thermal equilibrium as a consequence of its passivity
and robustness. 
In Sec. VI we take the limit of a dense bath and obtain,
by means of an analytic continuation method, a complex frequency $z_0=\Omega
+\delta \Omega +i\Gamma /2$ into which the unperturbed real frequency $%
\Omega $ of the oscillator is shifted by the heat bath. For a long period
of time the exponential decay law dominates the evolution. In this period the standard form 
of the Langevin
equation is derived, where the mean displacement of the Brownian
particle undergoes a slowly damped harmonic oscillation corresponding to the
complex frequency $z_0$.
Deviations from the exponential decay law are also discussed.\ 

\medskip
\section{Brownian motion: the model and its exact solution}
\medskip

Let us consider a harmonic oscillator interacting with a bath modeled by a
set of harmonic oscillators. The Hamiltonian of the system is

\begin{equation}
H=\frac{P^2}{2M}+\frac 12M\Omega ^2X^2+\sum\limits_{n=1}^N\left( \frac{p_n^2%
}{2m_n}+\frac 12m_n\omega _n^2x_n^2\right) +H_I,  \label{ho}
\end{equation}
where $H_I$ represents the interaction. Capital and lower-case letters 
stand for subsystem and bath variables respectively. In our case $H_I$ only involves a
linear coupling between the Brownian particle and the bath, i.e.

\begin{equation}
H_I=\sum\limits_{n=1}^Nc_n\left( Xx_n+\frac{Pp_n}{M\Omega m_n\omega _n}%
\right) ,  \label{hi}
\end{equation}
where all $c_n$ are real and small coupling constants. We define, as usual,
creation and annihilation operators ($\hbar =1)$

\begin{equation}
\begin{array}{c}
B=\sqrt{\frac{M\Omega }2}X+i\sqrt{\frac 1{2M\Omega }}P, \\ 
\\ 
b_n=\sqrt{\frac{m_n\omega _n}2}x_n+i\sqrt{\frac 1{2m_n\omega _n}}p_n,
\end{array}
\label{ab}
\end{equation}
that satisfy the canonical commutation relations 
\begin{equation}
\begin{array}{c}
\left[ B,B^{\dagger }\right] =I, \\ 
\\ 
\left[ b_n,b_m^{\dagger }\right] =\delta _{nm},
\end{array}
\label{rc}
\end{equation}
the other commutators vanish. In terms of these operators the Hamiltonian
reads

\begin{equation}
H=\Omega \left( B^{\dagger }B+\frac 12 \right) +\sum\limits_{n=1}^N
\omega _n \left( b_n^{\dagger }b_n+\frac 12 \right) +\sum\limits_{n=1}^Ng_n
\left(Bb_n^{\dagger }+B^{\dagger }b_n\right) ,  \label{hab}
\end{equation}
where $g_n=c_n/\sqrt{M\Omega m_n\omega _n}$. The linear interaction
allows us to find normal modes of $H$ in an exact way. This kind
of coupling is known in the literature as the rotating wave approximation 
\cite{West} (in general only the coupling between coordinates is taken into
account). 

In this model the interaction term preserves the total number of
quanta. In fact, defining the number of quanta operators

\[
\begin{array}{c}
N_\Omega =B^{\dagger }B, \\ 
\\ 
N_n=b_n^{\dagger }b_n,
\end{array}
\]
the total number of quanta given by

\[
N_T=N_\Omega +\sum\limits_{n=1}^NN_n 
\]
is a constant of motion, due to $dN_T/dt=i\left[ H_I,N_T\right] =0.$ Then we
can resolve the Hamiltonian into sectors of definite number of quanta. For $%
N_T=1$ (one-particle sector) and calling

\begin{eqnarray}
\left| \Omega \right\rangle &\equiv &B^{\dagger }\left| 0\right\rangle
=\left| 1\right\rangle \otimes \left| 0...0\right\rangle ,  \nonumber \\
&&  \label{ops} \\
\left| \omega _n\right\rangle &\equiv &b_n^{\dagger }\left| 0\right\rangle
=\left| 0\right\rangle \otimes \left| 0...%
%TCIMACRO{\QTATOPD. . {1}{n-{\rm site}} }
%BeginExpansion
{\textstyle {1 \atopwithdelims.. n-{\rm site}}}
%EndExpansion
...0\right\rangle ,  \nonumber
\end{eqnarray}
we obtain

\begin{equation}
H_1=\Omega \left| \Omega \right\rangle \left\langle
\Omega \right| +\sum\limits_{n=1}^N \omega _n \left|
\omega _n\right\rangle \left\langle \omega _n\right|
+\sum\limits_{n=1}^Ng_n\left( \left| \Omega \right\rangle \left\langle
\omega _n\right| +\left| \omega _n\right\rangle \left\langle \Omega \right|
\right)+C ,  \label{fri}
\end{equation}
where $C=\frac{\Omega}{2}+\sum\limits_{n=1}^N\frac{\omega_n}{2}$. This is the discrete version of the Friedrichs model \cite{Friedrichs}.

Let us find now the normal modes of $H,$ i.e. the new set of uncoupled
harmonic oscillators with normal frequencies $\alpha _\nu $ (Greek subscripts 
run from $0$ to $N$, while Arabic ones 
run from $1$ to $N).$ Hence we write $H$ as

\begin{equation}
H=\sum\limits_{\nu =0}^N\alpha _\nu c_\nu ^{\dagger }c_\nu +C ,  \label{hmn}
\end{equation}
where the new creation operators, $c_\nu ,$ are related to the old ones by
means of a unitary (canonical) transformation

\begin{equation}
c_\nu =\Phi _\nu B+\sum\limits_{n=1}^N\phi _{\nu n}b_n,  \label{tl}
\end{equation}
which preserves the canonical commutation relations

\begin{equation}
\left[ c_\mu ,c_\nu ^{\dagger }\right] =\delta _{\mu \nu }.  \label{cc}
\end{equation}

Coefficients

\begin{eqnarray*}
\Phi _\nu &=&\left\langle \alpha _\nu |\Omega \right\rangle , \\
&& \\
\phi _{\nu n} &=&\left\langle \alpha _\nu |\omega _n\right\rangle ,
\end{eqnarray*}
are the matrix elements of the unitary change of basis, from 
\{$\left|
\alpha _\nu \right\rangle $\} to
$\left\{ \left|
\Omega \right\rangle ,\left| \omega _n\right\rangle \right\}$, where

\[
\left| \alpha _\nu \right\rangle =c_\nu ^{\dagger }\left| 0\right\rangle 
\]
is an eigenvector of $H_1$

\[
H_1=\sum\limits_{\nu =0}^N\alpha _\nu \left| \alpha _\nu \right\rangle
\left\langle \alpha _\nu \right|+C . 
\]

The canonical commutators (\ref{cc}) impose the following condition for $%
\Phi _\nu $ and $\phi _{\nu n}$

\begin{equation}
\Phi _\mu \Phi _\nu ^{*}+\sum\limits_{n=1}^N\phi _{\mu n}\phi _{\nu
n}^{*}=\delta _{\mu \nu },  \label{co}
\end{equation}
which in the one-particle sector is a consequence of the orthogonality among
eigenvectors of $H_1,$

\[
\left\langle \alpha _\mu |\alpha _\nu \right\rangle =\delta _{\mu \nu }. 
\]

By taking into account transformation (\ref{tl}) in the Heisenberg equation
of motion for $c_\nu ,$%
\begin{equation}
i\frac{dc_\nu }{dt}=\left[ c_\nu ,H\right] =\alpha _\nu c_\nu ,  \label{ehc}
\end{equation}
and using 
\[
\begin{array}{c}
i\frac{dB}{dt}=\left[ B,H\right] =\Omega B+\sum\limits_{n=1}^Ng_nb_n, \\ 
\\ 
i\frac{db_n}{dt}=\left[ b_n,H\right] =\omega _nb_n+g_nB,
\end{array}
\label{eh} 
\]
we obtain the following system of linear equations

\begin{equation}
\begin{array}{c}
\Omega \Phi _\nu +\sum\limits_{n=1}^Ng_n\phi _{\nu n}=\alpha _\nu \Phi _\nu ,
\\ 
\\ 
g_n\Phi _\nu +\omega _n\phi _{\nu n}=\alpha _\nu \phi _{\nu n}.
\end{array}
\label{sel}
\end{equation}
From the second of these equations we can obtain $\phi _{\nu n}$ as

\begin{equation}
\phi _{\nu n}=\frac{g_n\Phi _\nu }{\alpha _\nu -\omega _n},  \label{phi}
\end{equation}
which is valid only if $\alpha _\nu \neq \omega _n,$ $\forall \nu ,n.$
Replacing it into the first equation of (\ref{sel}) we have

\begin{equation}
\Phi _\nu \left( \alpha _\nu -\Omega -\sum\limits_{n=1}^N\frac{g_n^2}{\alpha
_\nu -\omega _n}\right) =0.  \label{psi}
\end{equation}
Since we are looking for non-trivial solutions the expression between
brackets must be identically zero. Then we have an equation for the normal
frequencies of the new set of harmonic oscillators

\begin{equation}
\alpha _\nu -\Omega =\sum\limits_{n=1}^N\frac{g_n^2}{\alpha _\nu -\omega _n}.
\label{ea}
\end{equation}
The procedure developed above is equivalent to solve the eigenvalue problem $%
H_1\left| \alpha _\nu \right\rangle =\alpha _\nu \left| \alpha _\nu
\right\rangle $ for the matrix that represents the Friedrichs Hamiltonian in
the basis $\left\{ \left| \Omega \right\rangle ,\left| \omega
_n\right\rangle \right\} $

\begin{equation}
\left( 
\begin{array}{cccc}
\Omega & g_1 & 
\begin{array}{ccc}
\cdot & \cdot & \cdot
\end{array}
& g_N \\ 
g_1 & \omega _1 & 
\begin{array}{ccc}
&  & 
\end{array}
&  \\ 
\begin{array}{c}
\cdot \\ 
\cdot \\ 
\cdot
\end{array}
& 
\begin{array}{c}
\\ 
\\ 
\end{array}
& 
\begin{array}{ccc}
\cdot &  & 0 \\ 
& \cdot &  \\ 
0 &  & \cdot
\end{array}
& 
\begin{array}{c}
\\ 
\\ 
\end{array}
\\ 
g_N &  & 
\begin{array}{ccc}
&  & 
\end{array}
& \omega _N
\end{array}
\right) \left( 
\begin{array}{c}
\Phi _\nu ^{*} \\ 
\phi _{\nu 1}^{*} \\ 
\begin{array}{c}
. \\ 
. \\ 
.
\end{array}
\\ 
\phi _{\nu N}^{*}
\end{array}
\right) =\alpha _\nu \left( 
\begin{array}{c}
\Phi _\nu ^{*} \\ 
\phi _{\nu 1}^{*} \\ 
\begin{array}{c}
. \\ 
. \\ 
.
\end{array}
\\ 
\phi _{\nu N}^{*}
\end{array}
\right) .  \label{mat}
\end{equation}
In fact (\ref{mat}) is the complex conjugated matrix of the matrix form of
Eqs. (\ref{sel}) because $g_n$ is real. From Eq. (\ref{co}) for $\mu =\nu $
[normalization of the eigenvectors of (\ref{mat})] we obtain

\begin{equation}
\left| \Phi _\nu \right| ^2=\frac 1{1+\sum\limits_{n=1}^N\left( \frac{g_n}{%
\alpha _\nu -\omega _n}\right) ^2},  \label{nor}
\end{equation}
which can be completely determined if we know the set of eigenvalues $\alpha
_\nu .$ The normal frequencies $\alpha _\nu $ can be obtained numerically or
by analytic perturbative methods (in some special cases can even be obtained
exactly), so we assume that they are well known. In Fig. 1 we show where
these values are located.

\epsfysize=10truecm
\centerline{\epsffile{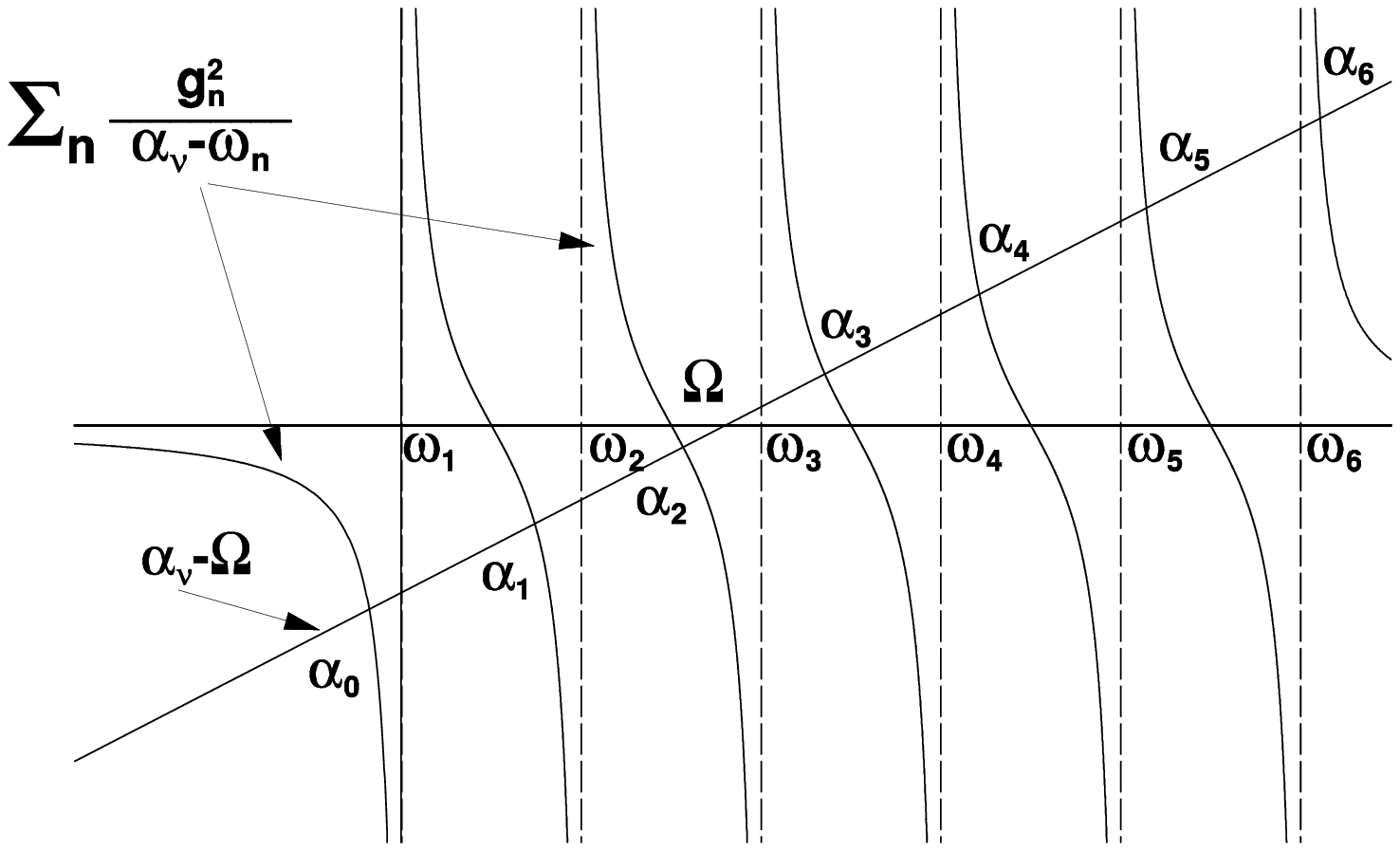}}
\centerline{{\bf FIG. 1.}\ Location of the normal frequencies.}
\bigskip
\bigskip

The normal frequencies correspond to the intersection of the straight line $%
\alpha _\nu -\Omega $ and the summation of the hyperboles $g_n^2/\left(
\alpha _\nu -\omega _n\right) .$ From this picture we see that the normal
frequencies always lie between consecutive frequencies of the unperturbed
Hamiltonian, except for the two extremum values which lie outside the
interval delimited by $\omega _1$ and $\omega _N$. \{$\alpha _\nu \}_{\nu
=0,...,N}$ never coincide with \{$\omega _n\}_{n=1,...,N},$ and are very
close to each $\omega _n$ near the extrema and move away from $\omega _n$ in
the centrum. In order to guarantee the positivity of the Hamiltonian as the
lowest frequency approaches zero and the convergence of the series $%
\sum\limits_{n=1}^N\frac{g_n^2}{\alpha _\nu -\omega _n},$ it is required that

\begin{equation}
\sum\limits_{n=1}^N\frac{g_n^2}{\omega _n-\omega _1+\delta }<\Omega -\omega
_1+\delta \hspace{0.3in}{\rm and}\hspace{0.3in}\sum\limits_{n=1}^N\frac{g_n^2%
}{\omega _N+\delta -\omega _n}<\omega _N+\delta -\Omega ,  \label{cp}
\end{equation}
where $\delta $ is an infinitesimal parameter (e.g. the distance between
contiguous unperturbed frequencies), and also that $g(\omega )$
behaves smoothly around $\omega _1$ and $\omega _N,$ having a small
value at these points. Then the coupling privileges the interaction with the
subsystem oscillator of frequency $\Omega .$ These conditions express the
fact that the interaction is small. However, for small $N$, 
the interaction must
not decrease very fast from the centrum since if such were the case the
subsystem oscillator would be coupled to few bath oscillators and then the 
bath would not be effective. These problems disappear approaching to the
continuum (Sec. VI).

\medskip
\section{Time evolution}
\medskip

We are looking for the way in which the relevant variables of the system
evolve in time. That is to know the time evolution of $B$ and 
$b_n$ and, from them, all other related dynamical variables. Coming back to
Eq. (\ref{ehc}) we can integrate it and obtain the temporal evolution of $%
c_\nu $

\begin{equation}
c_\nu (t)=e^{-i\alpha _\nu t}c_\nu (0).  \label{ct}
\end{equation}
Using the closure relation

\[
\sum\limits_{\nu =0}^N\left| \alpha _\nu \right\rangle \left\langle \alpha
_\nu \right| =I 
\]
we can obtain the following identities:

\begin{equation}
\begin{array}{c}
\sum\limits_{\nu =0}^N\Phi _\nu ^{*}\Phi _\nu =1, \\ 
\sum\limits_{\nu =0}^N\Phi _\nu ^{*}\phi _{\nu n}=0, \\ 
\sum\limits_{\nu =0}^N\phi _{\nu n}^{*}\phi _{\nu m}=\delta _{nm}.
\end{array}
\label{closrel}
\end{equation}
Eqs. (\ref{closrel}) allow us to perform the inverse of the transformation (%
\ref{tl}),

\begin{equation}
B=\sum\limits_{\nu =0}^N\Phi _\nu ^{*}c_\nu ,  \label{tia}
\end{equation}

\begin{equation}
b_n=\sum\limits_{\nu =0}^N\phi _{\nu n}^{*}c_\nu .  \label{tib}
\end{equation}
We are interested in knowing the explicit form of $B(t)$ and $b_n(t).$ Then,
from Eqs. (\ref{ct}), (\ref{tia}), (\ref{tib}), and (\ref{tl}) we have

\begin{equation}
\begin{array}{c}
B(t)=\sum\limits_{\nu =0}^N\Phi _\nu ^{*}e^{-i\alpha _\nu t}c_\nu
(0)=\sum\limits_{\nu =0}^N\Phi _\nu ^{*}e^{-i\alpha _\nu t}\left[ \Phi _\nu
B(0)+\sum\limits_{n=1}^N\phi _{\nu n}b_n(0)\right] , \\ 
\\ 
b_n(t)=\sum\limits_{\nu =0}^N\phi _{\nu n}^{*}e^{-i\alpha _\nu t}c_\nu
(0)=\sum\limits_{\nu =0}^N\phi _{\nu n}^{*}e^{-i\alpha _\nu t}\left[ \Phi
_\nu B(0)+\sum\limits_{m=1}^N\phi _{\nu m}b_m(0)\right] .
\end{array}
\label{paso}
\end{equation}
We have expressed the time evolution of the unperturbed annihilation
operators only in terms of their initial values. By considering Eq. (\ref
{phi}) we write Eq. (\ref{paso}) as

\begin{equation}
\begin{array}{c}
B(t)=\sum\limits_{\nu =0}^N\left| \Phi _\nu \right| ^2e^{-i\alpha _\nu
t}\left[ B(0)+\sum\limits_{n=1}^N\frac{g_n}{\alpha _\nu -\omega _n}%
b_n(0)\right] , \\ 
\\ 
b_n(t)=\sum\limits_{\nu =0}^N\left| \Phi _\nu \right| ^2\frac{g_n}{\alpha
_\nu -\omega _n}e^{-i\alpha _\nu t}\left[ B(0)+\sum\limits_{m=1}^N\frac{g_m}{%
\alpha _\nu -\omega _m}b_m(0)\right] ,
\end{array}
\label{atbtm}
\end{equation}
and a similar set of equations for the Hermitian conjugate operators. Eqs. (%
\ref{atbtm}) are determined uniquely by knowing the eigenvalues $\alpha _\nu
.$

These are the exact solutions of our problem. From them we can obtain all
the relevant information under consideration. For example, one of the
relevant variables is the position of the subsystem oscillator, 
$X(t)=\frac{1}{\sqrt{2M\Omega}}\left[
B^{\dagger }(t)+B(t)\right],$ given by

\begin{eqnarray}
X(t) &=&\sum\limits_{\nu =0}^N\left| \Phi _\nu \right| ^2\left\{ \left[ \cos
\left( \alpha _\nu t\right) X(0)+\sin \left( \alpha _\nu t\right) \widetilde{%
P}(0)\right] \right.  \nonumber  \label{xtot} \\
&&  \label{xtot} \\
&&+\frac 1{\sqrt{M\Omega }}\left. \sum\limits_{m=1}^N\frac{g_m}{\alpha _\nu
-\omega _m}\left[ \sqrt{m_n\omega _n}\cos \left( \alpha _\nu t\right) x_n(0)+%
\frac{\sin \left( \alpha _\nu t\right) }{\sqrt{m_n\omega _n}}p_n(0)\right]
\right\} ,  \nonumber
\end{eqnarray}
where $\widetilde{P}\equiv \frac P{M\Omega }.$ Similarly, from $\widetilde{P}%
=i\sqrt{\frac 1{2M\Omega }}\left[ B^{\dagger }(t)-B(t)\right] $ we have

\begin{eqnarray}
\widetilde{P}(t) &=&\sum\limits_{\nu =0}^N\left| \Phi _\nu \right| ^2\left\{
\left[ -\sin \left( \alpha _\nu t\right) X(0)+\cos \left( \alpha _\nu
t\right) \widetilde{P}(0)\right] \right.  \nonumber  \label{ptot} \\
&&  \label{ptot} \\
&&+\frac 1{\sqrt{M\Omega }}\left. \sum\limits_{n=1}^N\frac{g_n}{\alpha _\nu
-\omega _n}\left[ -\sqrt{m_n\omega _n}\sin \left( \alpha _\nu t\right)
q_n(0)+\frac{\cos \left( \alpha _\nu t\right) }{\sqrt{m_n\omega _n}}%
p_n(0)\right] \right\} .  \nonumber
\end{eqnarray}

Another interesting magnitude is the occupation number of the oscillator
representing the Brownian particle (subsystem dynamics). We have

\begin{eqnarray}
\left( B^{\dagger }B\right) (t) &=&\sum\limits_{\mu ,\nu =0}^N\left| \Phi
_\mu \right| ^2\left| \Phi _\nu \right| ^2e^{i(\alpha _\mu -\alpha _\nu
)t}\left[ \left( B^{\dagger }B\right) (0)+\sum\limits_{n=1}^N\frac{g_n}{%
\alpha _\mu -\omega _n}\left( Bb_n^{\dagger }\right) (0)\right.  \nonumber \\
&&  \label{sf} \\
&&\ \ +\left. \sum\limits_{n=1}^N\frac{g_n}{\alpha _\nu -\omega _n}\left(
B^{\dagger }b_n\right) (0)+\sum\limits_{n,m=1}^N\frac{g_ng_m}{(\alpha _\mu
-\omega _n)(\alpha _\nu -\omega _m)}\left( b_n^{\dagger }b_m\right)
(0)\right] .  \nonumber
\end{eqnarray}

Let us first consider the case in which the set of harmonic
oscillators modeling the bath is in thermal equilibrium inside a big
reservoir and the Brownian oscillator is isolated from the rest. At $t=0$
we extract the bath from the reservoir at temperature $T$, 
and put it in contact with the subsystem oscillator, in such a way that
the bath becomes a thermal reservoir for the Brownian particle. In this
situation the initial state of the total system is represented by a
time-independent density matrix which is a direct (tensorial) product of the
matrices representing the isolated harmonic oscillator $\rho _B(0)$ and the
environment degrees of freedom $\rho _b(0)$ in thermal equilibrium at
temperature $T$

\begin{equation}
\rho (0)=\rho _B(0)\otimes \frac{e^{-\beta H_b}}{{\rm tr}_b\left\{ e^{-\beta
H_b}\right\} },  \label{rho}
\end{equation}
where $H_b=\sum\limits_{n=1}^N \omega _n \left( b_n^{\dagger
}b_n +\frac 12 \right)$ is the bath Hamiltonian and {\rm tr}$_b$ is the partial trace over the
reservoir. If it is the case we have no correlations among the initial
states of subsystem and reservoir. 
To obtain the time evolution of $%
\left\langle \left( B^{\dagger }B\right) (t)\right\rangle \equiv {\rm tr}%
\left\{ \rho (0)\left( B^{\dagger }B\right) (t)\right\}$, we need to specify 
the
initial values of Eq. (\ref{sf}) in the state $\rho$ of Eq. (\ref{rho}). 
They are given by 

\begin{equation}
\begin{array}{c}
\left\langle \left( B^{\dagger }B\right) (0)\right\rangle =\kappa , \\ 
\left\langle \left( Bb_n^{\dagger }\right) (0)\right\rangle =0=\left\langle
\left( B^{\dagger }b_n\right) (0)\right\rangle , \\ 
\left\langle \left( b_n^{\dagger }b_m\right) (0)\right\rangle =\left(
e^{\beta \omega _n}-1\right) ^{-1}\delta _{nm},
\end{array}
\label{ci}
\end{equation}
where $\kappa $ is the initial number of quanta in the subsystem oscillator. Therefore the subsystem dynamics is given by

\begin{eqnarray}
\left\langle \left( B^{\dagger }B\right) (t)\right\rangle &=&\sum\limits_{%
%TCIMACRO{\QATOPD. . {\mu ,\nu =0}{\mu >\nu } }
%BeginExpansion
{\mu ,\nu =0 \atopwithdelims.. \mu >\nu }
%EndExpansion
}^N2\left| \Phi _\mu \right| ^2\left| \Phi _\nu \right| ^2\cos \left[ \left(
\alpha _\mu -\alpha _\nu \right) t\right] \left[ \kappa +\sum\limits_{n=1}^N%
\frac{g_n^2}{(\alpha _\mu -\omega _n)(\alpha _\nu -\omega _n)}\frac 1{%
e^{\beta \omega _n}-1}\right]  \nonumber  \label{sfci} \\
&&  \label{sfci} \\
&&+\sum\limits_{\nu =0}^N\left| \Phi _\nu \right| ^4\left[ \kappa
+\sum\limits_{n=1}^N\left( \frac{g_n}{\alpha _\nu -\omega _n}\right) ^2\frac 
1{e^{\beta \omega _n}-1}\right] .  \nonumber
\end{eqnarray}

\smallskip\ 

In a similar way we obtain the mean value of the number of quanta operator
for the $n-$oscillator of the bath, i.e.

\begin{center}
\begin{eqnarray}
\left\langle \left( b_n^{\dagger }b_n\right) (t)\right\rangle
&=&\sum\limits_{%
%TCIMACRO{\QATOPD. . {\mu ,\nu =0}{\mu >\nu } }
%BeginExpansion
{\mu ,\nu =0 \atopwithdelims.. \mu >\nu }
%EndExpansion
}^N\frac{2\left| \Phi _\mu \right| ^2\left| \Phi _\nu \right| ^2g_n^2\cos
\left[ \left( \alpha _\mu -\alpha _\nu \right) t\right] }{\left( \alpha _\mu
-\omega _n\right) \left( \alpha _\nu -\omega _n\right) }\left[ \kappa
+\sum\limits_{m=1}^N\frac{g_m^2}{(\alpha _\mu -\omega _m)(\alpha _\nu
-\omega _m)}\frac 1{e^{\beta \omega _m}-1}\right]  \nonumber  \label{bnt} \\
&&  \label{bnt} \\
&&+\sum\limits_{\nu =0}^N\left| \Phi _\nu \right| ^4\frac{g_n^2}{\left(
\alpha _\nu -\omega _n\right) ^2}\left[ \kappa +\sum\limits_{m=1}^N\left( 
\frac{g_m}{\alpha _\nu -\omega _m}\right) ^2\frac 1{e^{\beta \omega _m}-1}%
\right] .  \nonumber
\end{eqnarray}
\end{center}

The expressions obtained
can be formally rewritten as

\begin{eqnarray}
\left\langle N_\Omega (t)\right\rangle &=&P_{\Omega \Omega }(t)\left\langle
N_\Omega (0)\right\rangle +\sum\limits_{n=1}^NP_{\Omega n}(t)\left\langle
N_n(0)\right\rangle ,  \nonumber  \label{edgar} \\
&&  \label{edgar} \\
\left\langle N_n(t)\right\rangle &=&P_{n\Omega }(t)\left\langle N_\Omega
(0)\right\rangle +\sum\limits_{m=1}^NP_{nm}(t)\left\langle
N_m(0)\right\rangle ,  \nonumber
\end{eqnarray}
where $\left\langle N_\Omega \right\rangle =\left\langle B^{\dagger
}B\right\rangle $ and $\left\langle N_n\right\rangle =\left\langle
b_n^{\dagger }b_n\right\rangle .$ In a forthcoming paper we will show that
this is a general result of this kind of models, which allows us to derive the
Pauli master equation. It can be proved that $P_{\Omega \Omega }$ and $%
P_{\Omega n}$ are respectively, the transition probability of the
one-particle state $\left| \Omega \right\rangle $ remaining unchanged
(survival probability) and the transition probability from the state $\left|
\omega_n \right\rangle $ to the state $\left| \Omega \right\rangle $. They
represent the probability that at time $t$ the contribution to the
oscillator occupation number comes from itself and from the bath,
respectively. $P_{n\Omega }$ and $P_{nm}$ are the probability that the $n-$%
bath occupation number has contribution from the oscillator and from the
bath, respectively. These probabilities satisfy the normalization condition

\[
P_{\Omega \Omega }+\sum\limits_{n=1}^NP_{\Omega n}=1,\hspace{0.3in}%
P_{n\Omega }+\sum\limits_{m=1}^NP_{nm}=1, 
\]
and are explicitly given by 

\newpage

\begin{eqnarray}
P_{\Omega \Omega }(t) &\equiv &\left| \left\langle \Omega \left|
e^{-iHt}\right| \Omega \right\rangle \right| ^2=2\sum\limits_{%
%TCIMACRO{\QATOPD. . {\mu ,\nu =0}{\mu >\nu } }
%BeginExpansion
{\mu ,\nu =0 \atopwithdelims.. \mu >\nu }
%EndExpansion
}^N\left| \Phi _\mu \right| ^2\left| \Phi _\nu \right| ^2\cos \left[ \left(
\alpha _\mu -\alpha _\nu \right) t\right] +\sum\limits_{\nu =0}^N\left| \Phi
_\nu \right| ^4,  \nonumber \\
&&  \nonumber \\
P_{\Omega n}(t) &=&P_{n\Omega }(t)\equiv \left| \left\langle \Omega \left|
e^{-iHt}\right| \omega _n\right\rangle \right| ^2=2\sum\limits_{%
%TCIMACRO{\QATOPD. . {\mu ,\nu =0}{\mu >\nu } }
%BeginExpansion
{\mu ,\nu =0 \atopwithdelims.. \mu >\nu }
%EndExpansion
}^N\left| \Phi _\mu \right| ^2\left| \Phi _\nu \right| ^2\frac{g_n^2\cos
\left[ \left( \alpha _\mu -\alpha _\nu \right) t\right] }{(\alpha _\mu
-\omega _n)(\alpha _\nu -\omega _n)}  \nonumber \\
&&+\sum\limits_{\nu =0}^N\left| \Phi _\nu \right| ^4\left( \frac{g_n}{\alpha
_\nu -\omega _n}\right) ^2,  \label{pronm} \\
&&  \nonumber \\
P_{nm}(t) &\equiv &\left| \left\langle \omega _n\left| e^{-iHt}\right|
\omega _m\right\rangle \right| ^2=2\sum\limits_{%
%TCIMACRO{\QATOPD. . {\mu ,\nu =0}{\mu >\nu } }
%BeginExpansion
{\mu ,\nu =0 \atopwithdelims.. \mu >\nu }
%EndExpansion
}^N\left| \Phi _\mu \right| ^2\left| \Phi _\nu \right| ^2\frac{%
g_n^2g_m^2\cos \left[ \left( \alpha _\mu -\alpha _\nu \right) t\right] }{%
(\alpha _\mu -\omega _n)(\alpha _\nu -\omega _n)(\alpha _\mu -\omega
_m)(\alpha _\nu -\omega _m)}  \nonumber \\
&&+\sum\limits_{\nu =0}^N\left| \Phi _\nu \right| ^4\left[ \frac{g_ng_m}{%
\left( \alpha _\nu -\omega _n\right) \left( \alpha _\nu -\omega _m\right) }%
\right] ^2.  \nonumber
\end{eqnarray}

We can see from the second equation of (\ref{edgar}) that, although there is
no interaction term in the Hamiltonian among the bath oscillators 
themselves, the time evolution for a bath oscillator has
contributions coming from the whole bath. This fact was noticed in Ref. \cite
{Gruver} [cf. Eqs. (2.2e) and (2.2f)].

The decomposition made in Eq. (\ref{edgar}) is useful for studying the
different contributions to the time evolution of the mean number operators 
(Sec. V).

In the limit of low temperatures $T\rightarrow 0$, $\left\langle
N_n(0)\right\rangle =\left( e^{\beta \omega _n}-1\right) ^{-1}\rightarrow 0,$
and then $\left\langle N_\Omega (t)\right\rangle =P_{\Omega \Omega
}(t)\left\langle N_\Omega (0)\right\rangle =\kappa P_{\Omega \Omega }(t).$
In the case $\kappa =1$ it is the survival probability 
(the probability of no decay of the state $\left| \Omega \right\rangle $), 

\begin{equation}
\left. \left\langle N_\Omega (t)\right\rangle \right| _{T=0}=\left|
\left\langle \Omega \left| e^{-iHt}\right| \Omega \right\rangle \right| ^2,%
\hspace{0.3in}{\rm for}\ \left\langle N_\Omega (0)\right\rangle =1.
\label{surp}
\end{equation}
In Sec. VI for a dense bath we show that the asymptotic behavior of this
probability obeys a power-law decay. This is a well known fact in decay theory
of unstable quantum systems and was reported as an anomaly in statistical treatments of quantum open
systems (see, e.g., Ref. \cite{Haake}).

In the next section we derive an exact equation of motion of the mean value
of the position operator $X$ (a generalized form of the Langevin equation).

\medskip
\section{Langevin equation\ }
\medskip

Let us consider Eqs. (\ref{xtot}) and (\ref{ptot}) for the bath in thermal
equilibrium at the initial time. In this case let $\{\left| N_n\right\rangle
\}$ be a basis of eigenvectors of $N_n.$ Taking mean values in
the state (\ref{rho}) we have

\[
\left\langle b_n(0)\right\rangle ={\rm tr}\left\{ b_n(0)\frac{\rho_B(0)\exp \left[
-\beta \omega _n\left( N_n+1/2\right) \right] 
\prod\limits_{i=1,i\neq n}^N\exp \left[ -\beta \omega _i\left(
N_i+1/2\right) \right] }{{\rm tr}_b\left\{ e^{-\beta H_b}\right\} }\right\} .
\]
There is a vanishing factor  $\sum_{N_n}\left\langle N_n\right|
b_n(0)\exp \left[ -\beta \omega _n\left( N_n+1/2\right) \right] \left|
N_n\right\rangle ,$ since $\left\langle N_n\right| b_n(0)\left|
N_n\right\rangle =0.$ Then $\left\langle b_n(0)\right\rangle =0$. 
Similarly $\left\langle b_n^{\dagger }(0)\right\rangle =0$. So we have $\left\langle
q_n(0)\right\rangle =0=\left\langle p_n(0)\right\rangle .$ Thus

\begin{eqnarray}
\left\langle X(t)\right\rangle &=&\sum\limits_{\nu =0}^N\left| \Phi _\nu
\right| ^2\left[ \cos \left( \alpha _\nu t\right) \left\langle
X(0)\right\rangle +\sin \left( \alpha _\nu t\right) \left\langle \widetilde{P%
}(0)\right\rangle \right] ,  \nonumber  \label{comi} \\
&&  \label{comi} \\
\left\langle \widetilde{P}(t)\right\rangle \ &=&\sum\limits_{\nu =0}^N\left|
\Phi _\nu \right| ^2\left[ -\sin \left( \alpha _\nu t\right) \left\langle
X(0)\right\rangle +\cos \left( \alpha _\nu t\right) \left\langle \widetilde{P%
}(0)\right\rangle \right] .  \nonumber
\end{eqnarray}
The initial and instantaneous variables are related by a generalized sum of
rotations

\begin{equation}
\left( 
\begin{array}{c}
\left\langle X(t)\right\rangle \\ 
\left\langle \widetilde{P}(t)\right\rangle
\end{array}
\right) =\sum\limits_{\nu =0}^N\left| \Phi _\nu \right| ^2\left( 
\begin{array}{cc}
\cos (\alpha _\nu t) & \sin (\alpha _\nu t) \\ 
-\sin (\alpha _\nu t) & \cos (\alpha _\nu t)
\end{array}
\right) \left( 
\begin{array}{c}
\left\langle X(0)\right\rangle \\ 
\left\langle \widetilde{P}(0)\right\rangle
\end{array}
\right) ,  \label{rotat}
\end{equation}
a transformation which can be summarized as

\begin{equation}
\left( 
\begin{array}{c}
\left\langle X(t)\right\rangle \\ 
\left\langle \widetilde{P}(t)\right\rangle
\end{array}
\right) =\left( 
\begin{array}{cc}
a(t) & b(t) \\ 
-b(t) & a(t)
\end{array}
\right) \left( 
\begin{array}{c}
\left\langle X(0)\right\rangle \\ 
\left\langle \widetilde{P}(0)\right\rangle
\end{array}
\right) ,  \label{qporp}
\end{equation}
where $a(t)=\sum\limits_{\nu =0}^N\left| \Phi _\nu \right| ^2\cos \left(
\alpha _\nu t\right) $ and $b(t)=\sum\limits_{\nu =0}^N\left| \Phi _\nu
\right| ^2\sin \left( \alpha _\nu t\right) .$

We can invert the matrix of Eq. (\ref{qporp}) to obtain $\left\langle
X(0)\right\rangle $ and $\left\langle \widetilde{P}(0)\right\rangle $ as
functions of $\left\langle X(t)\right\rangle $ and $\left\langle \widetilde{P%
}(t)\right\rangle $ :

\begin{equation}
\left( 
\begin{array}{c}
\left\langle X(0)\right\rangle \\ 
\left\langle \widetilde{P}(0)\right\rangle
\end{array}
\right) =\frac 1\Delta \left( 
\begin{array}{cc}
a(t) & -b(t) \\ 
b(t) & a(t)
\end{array}
\right) \left( 
\begin{array}{c}
\left\langle X(t)\right\rangle \\ 
\left\langle \widetilde{P}(t)\right\rangle
\end{array}
\right) ,  \label{trin}
\end{equation}
where $\Delta (t)=a^2(t)+b^2(t).$ By taking time derivatives in Eq. (\ref
{qporp}) and replacing the initial mean values by those of Eq. (\ref{trin})
we have

\begin{equation}
\left( 
\begin{array}{c}
\left\langle \stackrel{.}{X}(t)\right\rangle \\ 
\left\langle \stackrel{.}{\widetilde{P}}(t)\right\rangle
\end{array}
\right) =\frac 1\Delta \left( 
\begin{array}{cc}
\stackrel{.}{a}a+\stackrel{.}{b}b \ & \stackrel{.}{b}a-\stackrel{.}{a}b \\ 
\stackrel{.}{a}b-\stackrel{.}{b}a \ & \stackrel{.}{a}a+\stackrel{.}{b}b
\end{array}
\right) \left( 
\begin{array}{c}
\left\langle X(t)\right\rangle \\ 
\left\langle \widetilde{P}(t)\right\rangle
\end{array}
\right) .  \label{dpri}
\end{equation}
Similarly, from the second derivative of (\ref{qporp}) we have

\begin{equation}
\left( 
\begin{array}{c}
\left\langle \stackrel{..}{X}(t)\right\rangle \\ 
\left\langle \stackrel{..}{\widetilde{P}}(t)\right\rangle
\end{array}
\right) =\frac 1\Delta \left( 
\begin{array}{cc}
\stackrel{..}{a}a+\stackrel{..}{b}b \ & \stackrel{..}{b}a-\stackrel{..}{a}b \\ 
\stackrel{..}{a}b-\stackrel{..}{b}a \ & \stackrel{..}{a}a+\stackrel{..}{b}b
\end{array}
\right) \left( 
\begin{array}{c}
\left\langle X(t)\right\rangle \\ 
\left\langle \widetilde{P}(t)\right\rangle
\end{array}
\right) .  \label{dsec}
\end{equation}
Finally eliminating $\left\langle \widetilde{P}(t)\right\rangle $ from 
Eq. (\ref{dpri}) we obtain a generalized
form of the Langevin equation with time-dependent coefficients, 

\begin{equation}
\left\langle \stackrel{..}{X}(t)\right\rangle +\Omega ^2(t)\left\langle
X(t)\right\rangle +\Gamma (t)\left\langle \stackrel{.}{X}(t)\right\rangle =0,
\label{eclan}
\end{equation}
where

\begin{equation}
\Omega ^2(t)=\frac{\stackrel{.}{a}\ \stackrel{..}{b}-\stackrel{.}{b}%
\ \stackrel{..}{a}}{a\stackrel{.}{b}-b\stackrel{.}{a}},\hspace{1.0in}%
\Gamma (t)=\frac{b\stackrel{..}{a}-a\stackrel{..}{b}}{a\stackrel{.}{b}-b%
\stackrel{.}{a}}.  \label{coeclan}
\end{equation}
The standard
stochastic force $f_{{\rm stoch}}$ does not appear in Eq. (\ref{eclan})
since it is included in the terms containing the operators $q_n(0)$ and $%
p_n(0)$, which were eliminated by taking the mean values in a thermal
equilibrium initial state of the bath. That is $\left\langle f_{{\rm stoch}%
}(t)\right\rangle =0.$ Eq. (\ref{eclan}) contrasts with the equivalent, but
non-local in time, standard integro-differential form of the equation of
motion of $\left\langle X\right\rangle.$ Eq. (\ref{eclan}) is actually a rather complicated expression
since the coefficients are not easy of evaluating. In Sec. VI we estimate
them in the continuous limit.

\medskip
\section{Examples and results}
\medskip

Let us describe the model we used for obtaining the numerical results of
this section.

\smallskip
\subsection{Choice of parameters}
\medskip

The model described in Secs. II and III consists of three main
ingredients: the subsystem and the bath, the interaction, and the initial
conditions. We have considered a subsystem represented by a harmonic
oscillator with natural frequency $\Omega $ and mass $M$ (heavy Brownian
particle), a bath of small oscillators with frequencies $\omega _n$ varying
in a range between $\omega _{\min }$ and $\omega _{\max },$ a small linear
coupling between system and bath, $g_n=\lambda c_n^{\prime },$ where $%
\lambda =(M\Omega )^{-1/2}$ and $c_n^{\prime }=c_n\left( m_n\omega _n\right)
^{-1/2},$ and the whole composed system prepared in such a way that at $%
t=0^{-}$ there is no correlation between subsystem and bath. The bath is
in equilibrium with an external heat source at temperature $T=\left(
k_B\beta \right) ^{-1},$ where $k_B$ is the Boltzmann constant, and at $%
t=0^{+}$ the bath is extracted from the thermal source, put in contact with
the subsystem and the total system is left isolated. Each parameter
mentioned above defines a typical time scale. These are: the scale associated with the
natural frequency of the isolated subsystem, $\Omega ^{-1};$ the scale
defined by the lowest frequency of the bath, $\omega _{\min }^{-1},$ related
to the reaction of the system when the interaction is switched on; the decay
time $\Gamma ^{-1}$ [see Eq. (69) in Sec. VI] in which the subsystem
dissipates its energy into the reservoir, and which is related to the squared of the
perturbation parameter $\lambda ;$ the memory time related with the highest
frequency present in the bath, $\omega _{\max }^{-1};$ the time scale $\beta$ 
associated with thermal effects (relative to quantum ones); the
Poincar\'e recurrence time given by the minimal difference between 
contiguous normal frequencies, specifically $t_P\simeq \frac{2\pi }{%
\min \left( \alpha _{\nu +1}-\alpha _\nu \right) }$ [see Eq. (48) below];
and two time scales related with quantum deviations form the exponential
decay law, a very short time $t_Z$ (Zeno period, which is responsible for
no decaying of the subsystem under a continuous succession of measurements
and occurs because of the temporal derivative of the survival amplitude
vanishes at $t=0$) and a very long one $t_K$ (Khalfin period of power
series tails, which is a consequence of the lower bound of the energy) (see
Sec. VI). In order to have a manifestation of these time scales the
parameters and variables must be chosen with certain criterion. An important
condition we must take into account and which is frequently overlooked in
the literature is condition (\ref{cp}). For example, 
in the case of a semi-infinite frequency spectrum, $%
\omega \in (0,\infty ),$ 
the often used
ohmic spectral density, $g^2(\omega )\sim \omega ,$ does not satisfy
condition (\ref{cp}), 
since it has a logarithmic type divergency. 
As we want to obtain a pictorial image
of the temporal evolution of the main magnitudes of Sec. III, then we
specify the parameters appearing in these magnitudes [e.g. Eqs. (\ref{xtot}%
), (\ref{ptot}), (\ref{sfci}), and (\ref{bnt})] as follows:

\begin{eqnarray}
\Omega &=&1,  \nonumber  \label{param} \\
\beta &=&\frac 1\Omega ,  \label{param} \\
\kappa &=&1.  \nonumber
\end{eqnarray}
This choice of $\beta $ fixes the thermal time scale to the same value of the 
$\Omega ^{-1}$ scale, and then purely quantum-mechanical and thermal effects
are comparable, so we are far of the classical limit $\hbar \Omega \ll k_BT$%
. The choice of $\kappa $ facilitates the comprehension of the one-particle
sector several times studied in decay theory.

For the sake of simplicity we consider for the variables  the case in which
the bath frequencies are equidistant around the frequency $\Omega ,$
i.e.

\begin{equation}
\omega _n=\Omega +A\left( n-\frac{N+1}2\right) ,\hspace{0.3in}\ 
n=1,...,N,  \label{osn}
\end{equation}
where $A$ is  
the spacing between 
contiguous frequencies of the bath, $A=\omega _{n+1}-\omega _n,$ 
being the band width $\omega_N -\omega_1 = A(N-2)$, and the 
number of small oscillators $N$ is an odd integer. 

The coupling function is given by a Lorentzian-like function

\begin{equation}
g_n=\frac{Da^2}{a^2+\left( \omega _n-\Omega \right) ^2}.  \label{gsn}
\end{equation}
This function is plotted in Fig. 2. $D$ is related to the coupling
strength $\lambda $ and is taken equal to $A,$ 
for reasons
which will become clear below. We fix $a=\frac{A(N-2)}2$ in order
to have half of the maximum value of $g_n$ at the
extrema. Finally we take the band width 
equal to $0.018,$ for all $N,$ which is the value that
allows us to compare our numerical results with those obtained by Gruver 
{\it et al. }\cite{Gruver} (who solved a set of coupled differential
equations coming from a maximum entropy principle approach) in the case $%
N+1=32.$ In their work the choice of the coupling function is different
and also different the criterion to increase the number of small
oscillators. While we maintain fixed the band width, they maintain fixed the value of the spacing $A.$
As one of the purposes of this work is to study the way of reaching the
continuous limit (see Sec. VI), 
we must have $A\rightarrow 0,$ $N\rightarrow \infty ,$ and $AN=$const.

\epsfysize=5truecm
\epsfysize=9truecm
\centerline{\epsffile{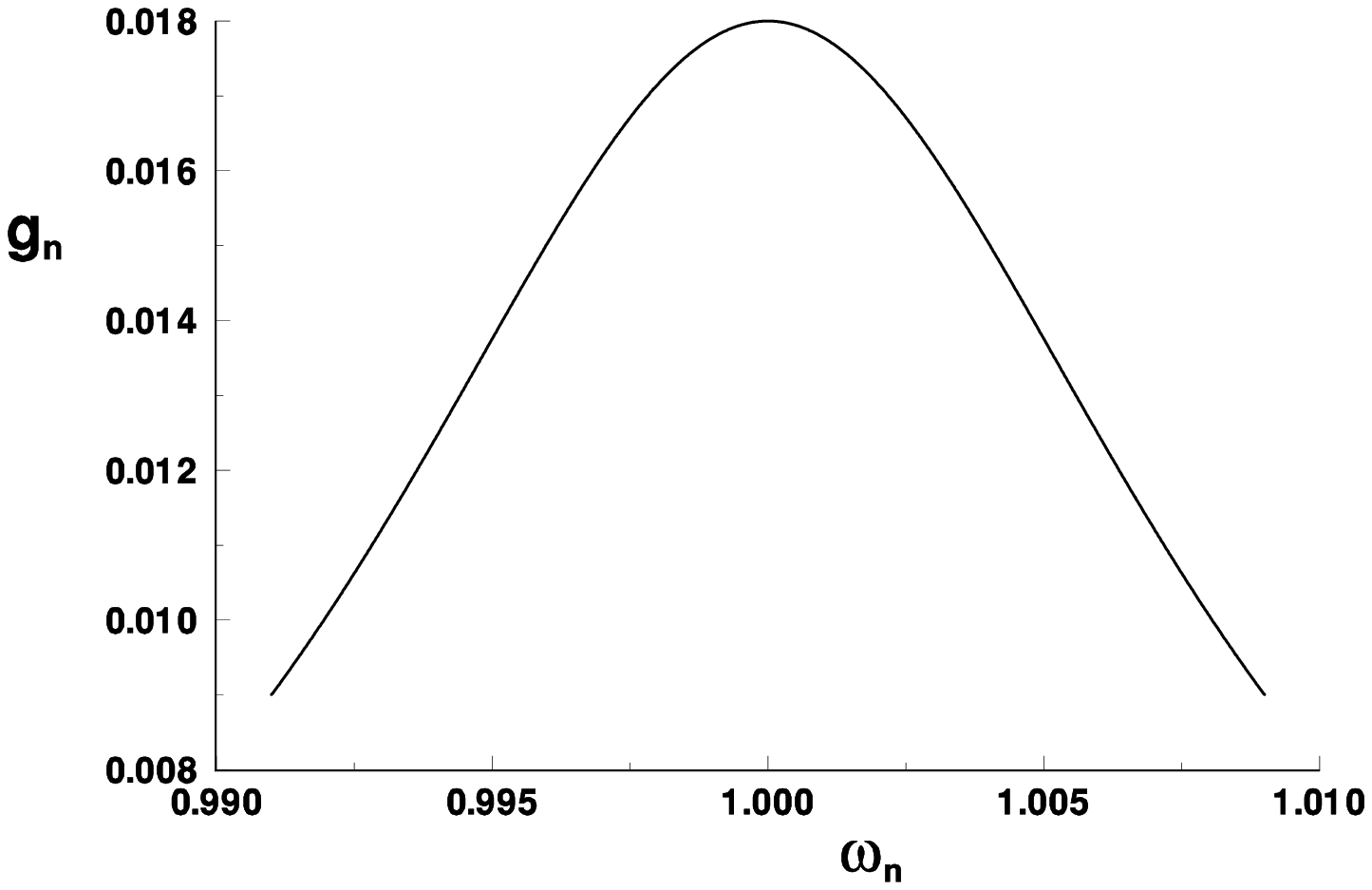}}
\centerline{{\bf FIG. 2.}\ The coupling function.}
\bigskip
\bigskip

Let us see the criterion used for fixing $D.$ It is a consequence of
conditions (\ref{cp}). Taking $\delta =A$ we have

\[
\sum\limits_{n=1}^N\frac{g_n^2}{\omega _N+A-\omega _n}<\omega _N+A-\Omega . 
\]
From Eq. (\ref{osn}) we have $\omega _N+A-\Omega =\frac{A(N+1)}2,$ $\omega
_N+A-\omega _n=A(N+1-n)$, and for a large bath we can approximate $g_n^2\simeq 
\frac{D^2N^4}{16\left[ \left( N/2\right) ^2+\left( N/2-n\right) ^2\right] ^2}%
.$ Then it must be satisfied

\[
\frac{D^2N^3}{8A^2}\sum\limits_{n=1}^N\frac 1{(N+1-n)\left[ \left( \frac N2%
\right) ^2+\left( \frac N2-n\right) ^2\right] ^2}<1. 
\]
The summation can be bounded by $N$ times its maximum value, which is
reached for $n=N.$ Thus

\[
\frac{D^2N^3}{8A^2}N\frac 1{\left( \frac N2\right) ^2}=\frac{D^2}{2A^2}<1, 
\]
which implies

\begin{equation}
D<\sqrt{2}A.  \label{cota}
\end{equation}
So we find an upper boundary for $D.$ In general we choose $D=A.$

The $N+1$ normal frequencies $\alpha _\nu $ are obtained from a 
matrix-diagonalization routine. 

\smallskip
\subsection{Numerical results}
\medskip

In Figs. 3 to 7 we plot 
$\left\langle N_\Omega \right\rangle $ ${\it vs.}$ $\Omega t.
$ We show that $\left\langle N_\Omega \right\rangle $ decays in
time to an asymptotic value, when the subsystem reaches equilibrium with the
bath, given by $(e-1)^{-1}\simeq 0.582$ [see Eq. (77) in Sec. VI, recall
$\beta \Omega =1$]. After a long period a revival appears reaching
again a similar value to the initial condition. Since for positive values of 
$t$ the arguments of the cosines in Eq. (\ref{sfci}) never are exactly
in phase, the revival does not fully reconstruct the initial condition, so that
the peak is smaller than the initial one and slightly broadens. Then, in the
continuous limit, the time of revival goes to infinity and the peak gets out
of sight among thermal fluctuations. As shown in Fig. 9, this revival is
periodic in time. It corresponds to the Poincar\'e recurrence time and is
given by the inverse of the smallest difference of normal frequencies, since
Eq. (\ref{sfci}) is a quasi-Fourier series because of the quasi-equidistance
of the normal frequencies. Specifically 

\begin{equation}
t_P\simeq \frac{2\pi }{\min \left( \alpha _{\nu +1}-\alpha _\nu \right) },
\label{tpoin}
\end{equation}
where obviously $\min \left( \alpha _\mu -\alpha _\nu \right) =\min \left(
\alpha _{\nu +1}-\alpha _\nu \right) .$ Table I shows this time for
different values of $N+1$.

%\smallskip\ 

\begin{center}
{\bf Tab. I}. Poincar\'e recurrence time.\\
\medskip\ 
\begin{tabular}{|c|c|c|c|c|}
\hline
$N+1$ &   $10$ &    $32$ &   $100$ &    $500$ \\ \hline
$t_P$ & $3370$ & $11190$ & $37311$ & $177994$ \\ \hline
\end{tabular}
\end{center}

Fig. 3 shows that for a small number of bath oscillators, it does
not result effective and loses the necessary robustness to break the
natural oscillations of the subsystem. For $N+1=10$ (Fig. 4) the energy lost 
of the subsystem oscillator is close to a dissipative behavior and
the bath begins to be effective. It is surprising that for so few bath
oscillators the subsystem already dissipates. With increasing $N$, fluctuations
get smaller and, since the spacing between frequencies decreases, $t_P$
grows. In Fig. 8 we draw $\left\langle N_\Omega \right\rangle $ for
different values of $N+1$ ${\it vs.}$ a re-scaled time with respect to $t_P$ $%
(\Omega =1).$ It shows the tendency to an exponential decay when the model
approaches to the continuum. Nevertheless in the continuous limit the
exponential decay law is not exact as we show in Sec. VI.

\vspace{0.5in}
\epsfysize=7.5truecm
\centerline{\epsffile{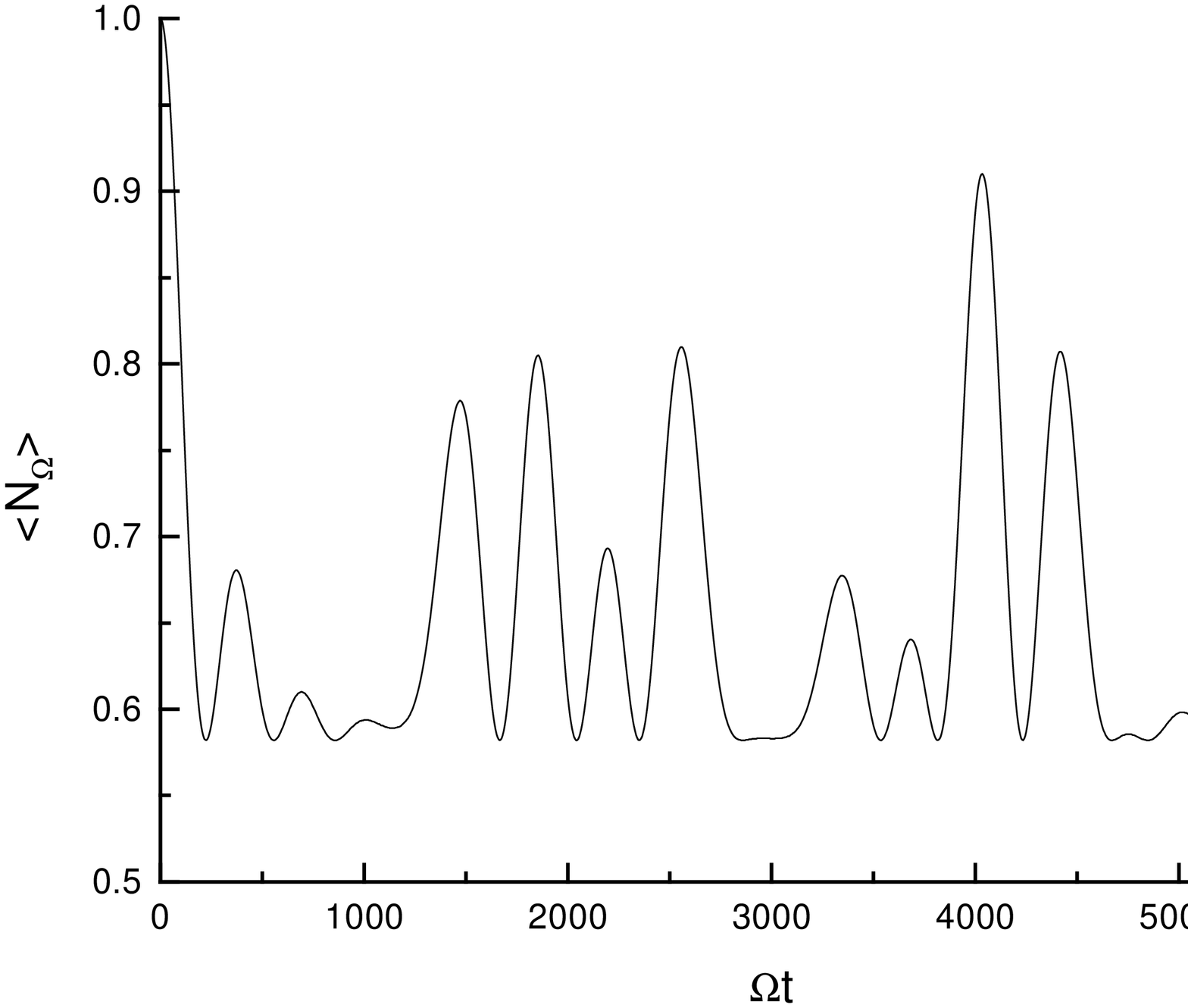}\hspace{1.15in}\epsfysize=7.5truecm\epsffile{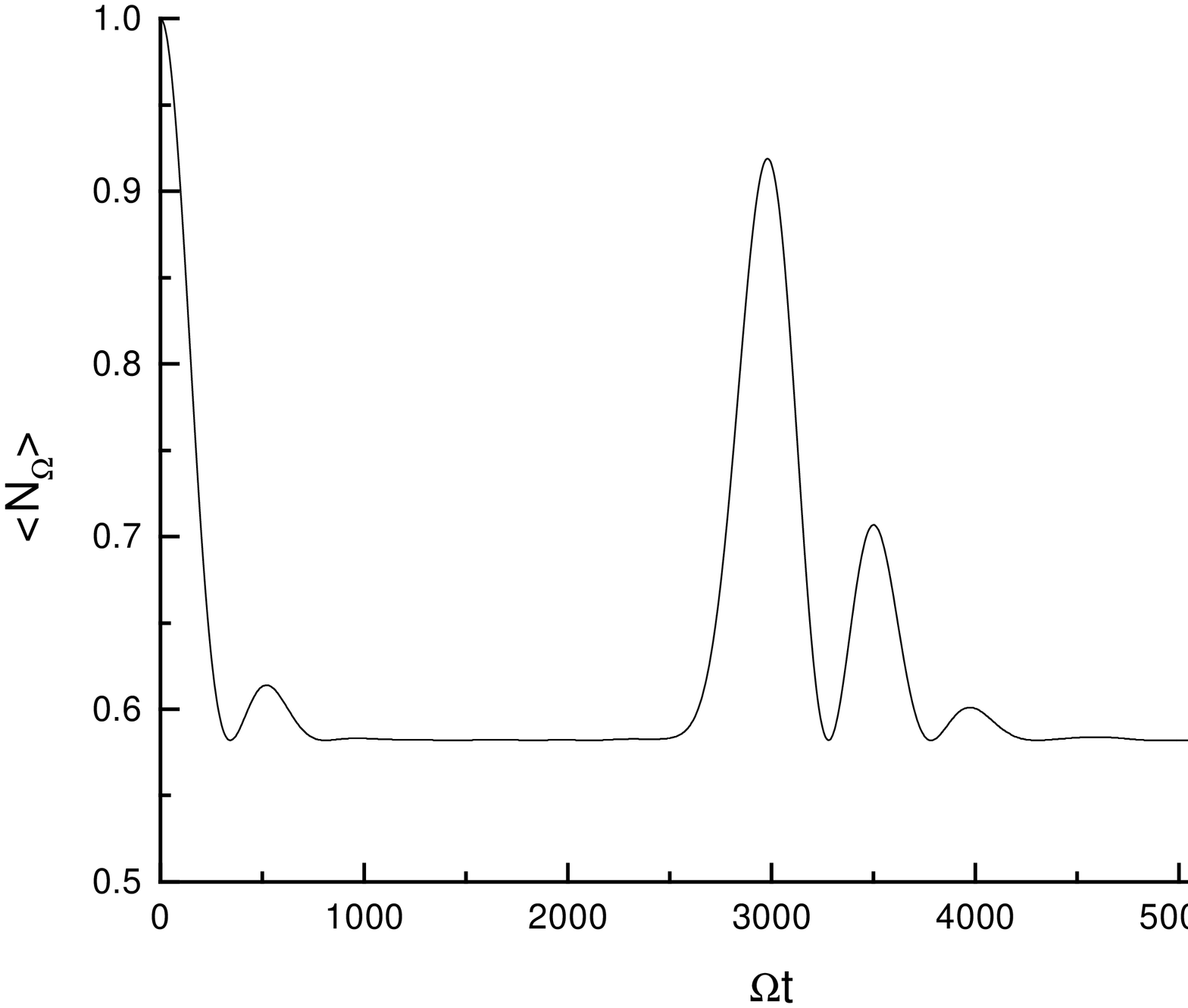}}
\centerline{{\bf FIG. 3.}\ $N+1=6$. \hspace{2.in} {\bf FIG. 4.}\ $N+1=10$.}
\vspace{0.6in}
\epsfysize=7.5truecm
\centerline{\epsffile{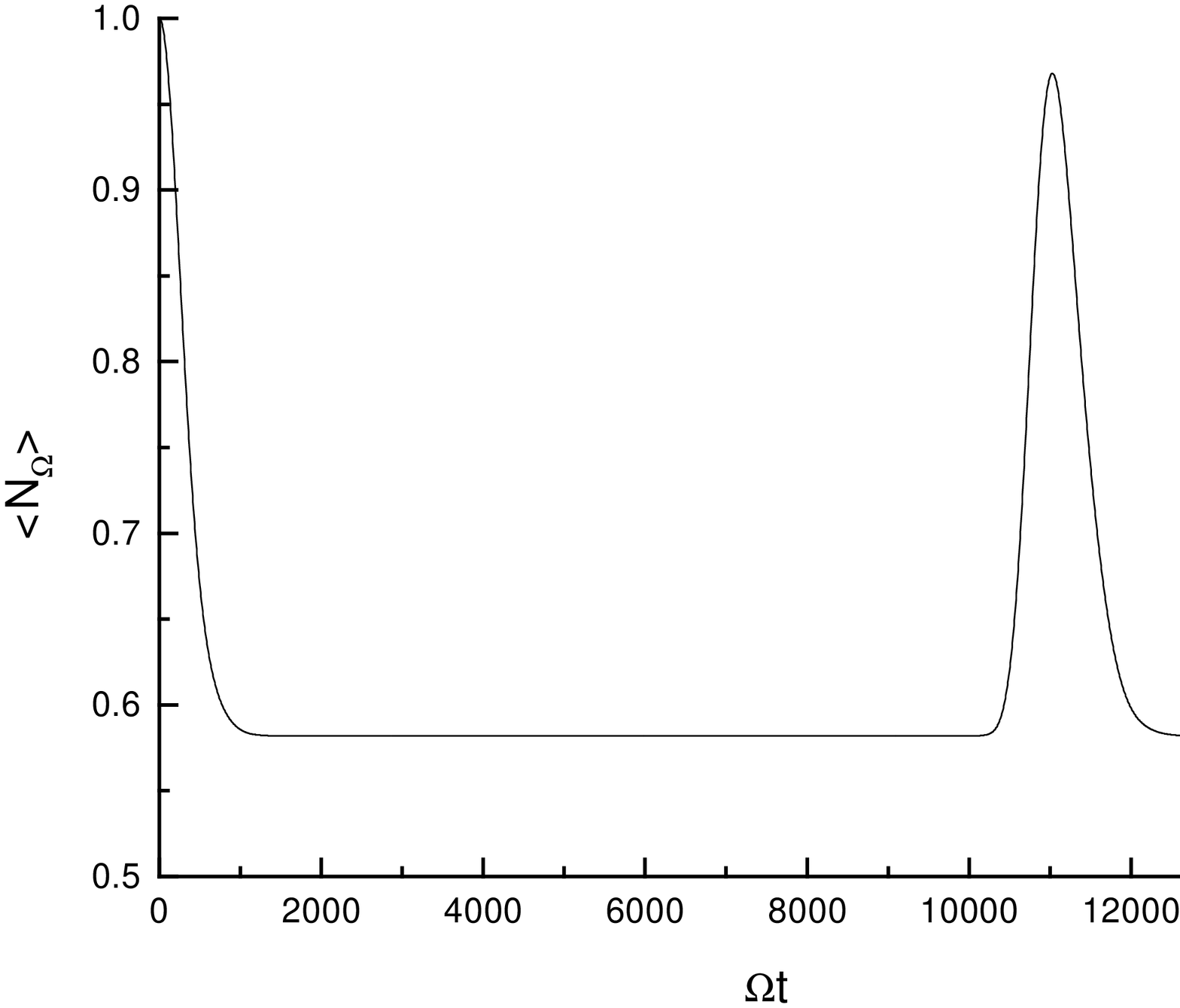}\hspace{1.15in}\epsfysize=7.5truecm\epsffile{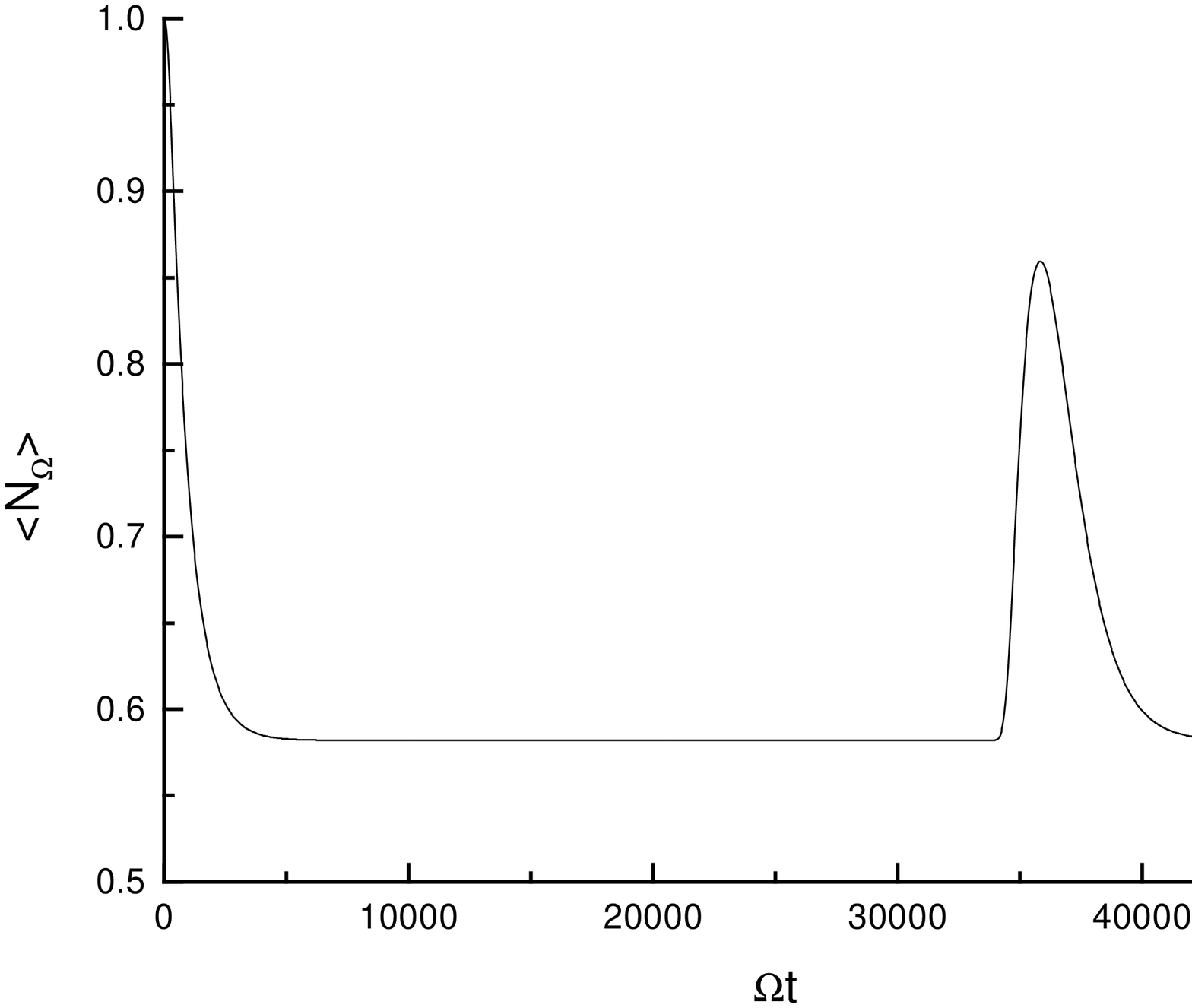}}
\centerline{{\bf FIG. 5.}\ $N+1=32$. \hspace{2.in} {\bf FIG. 6.}\ $N+1=100$.}
\vspace{0.6in}
\epsfysize=7.5truecm
\centerline{\epsffile{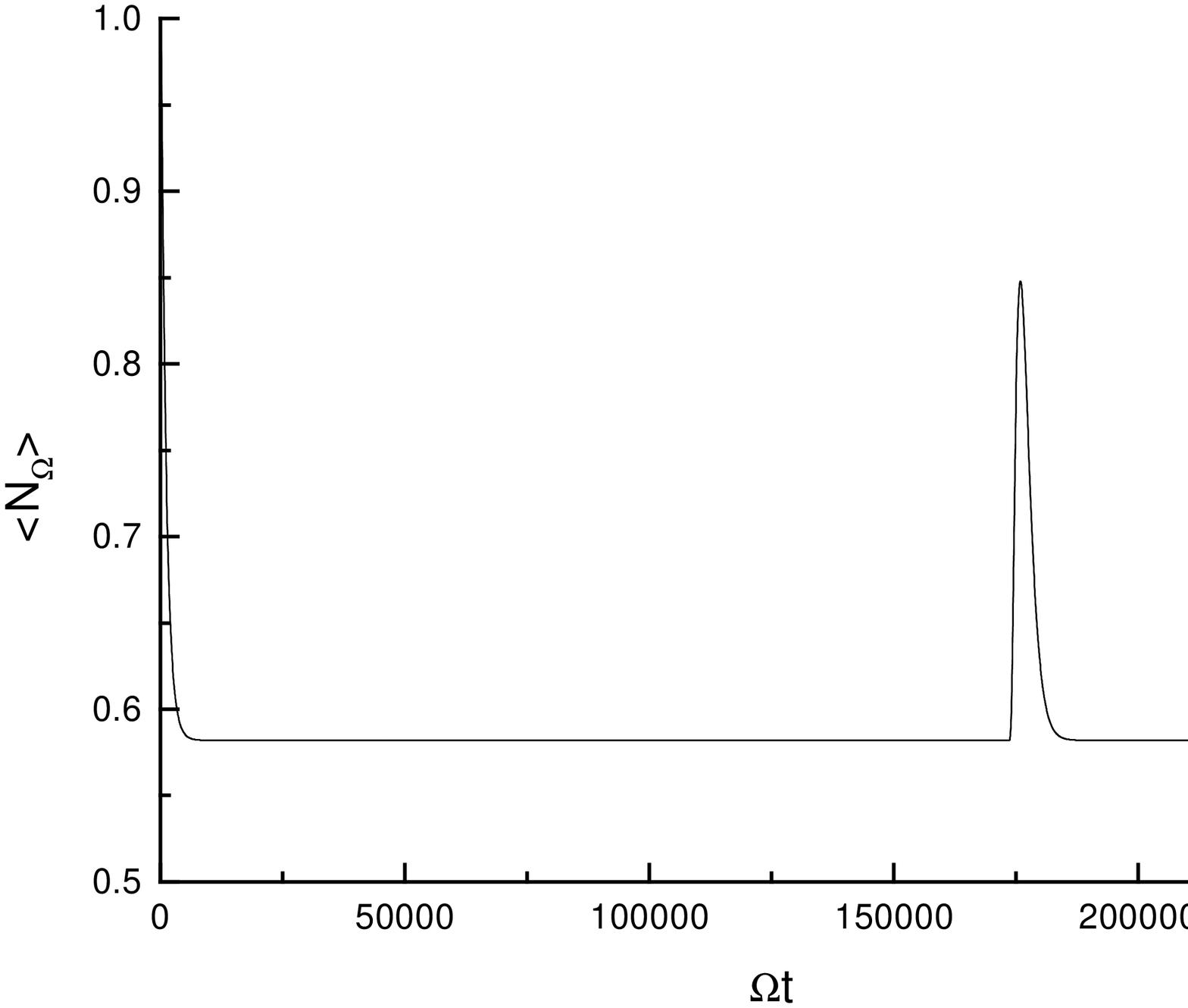}\hspace{1.15in}\epsfysize=7.5truecm\epsffile{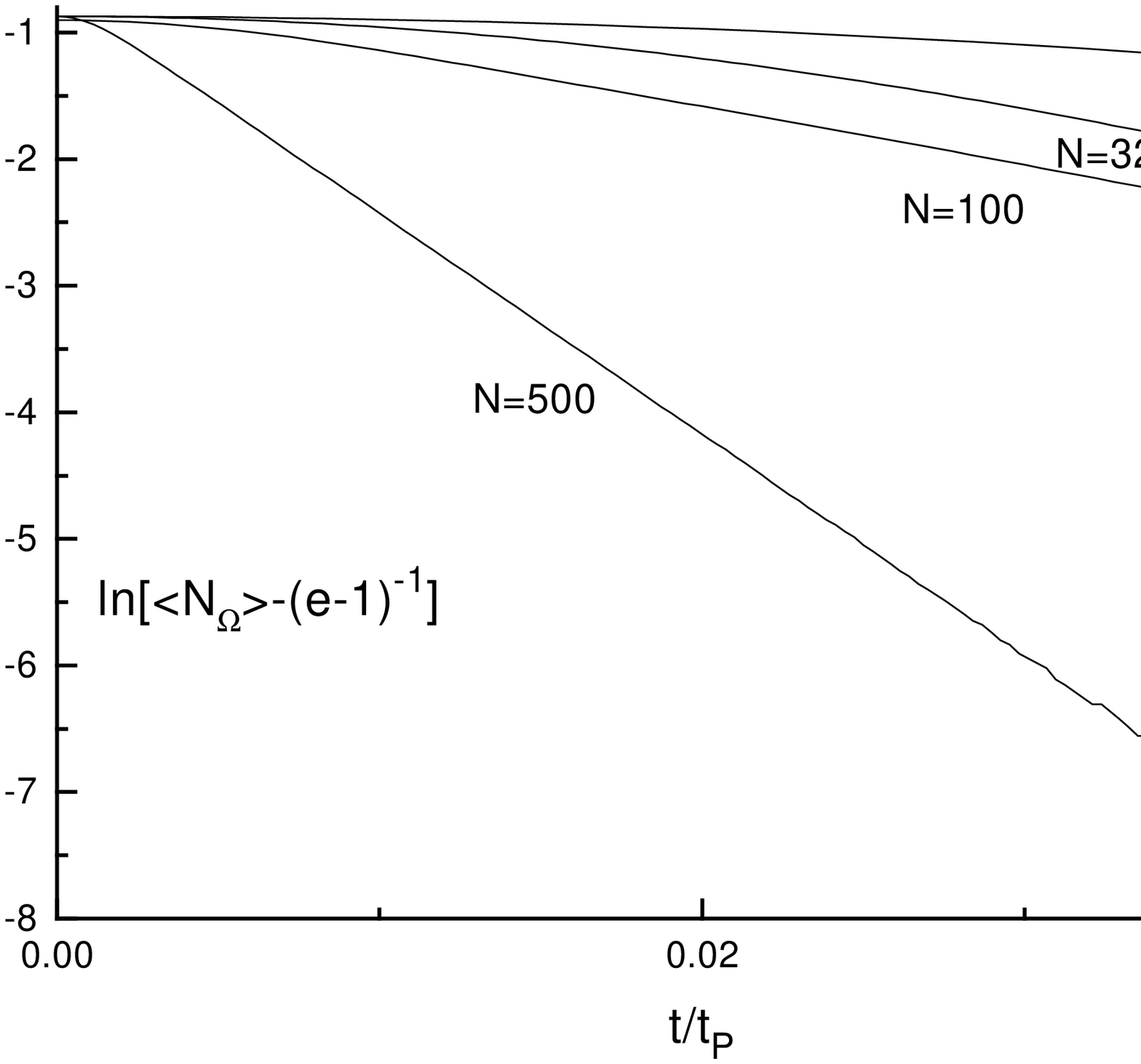}}
\centerline{\ \ \ \ \ \ \ \ \ {\bf  FIG.    7.}\  $N+1=500$.    \hspace{1.5in}  {\bf  FIG.    8.}\
Exponential approximation.}
\vspace{0.6in}

Figs. 9 and 10 show the behavior of $\left\langle N_\Omega \right\rangle $
for very long times in order to see the periodicity of this magnitude and
the recurrence of $t_P.$ In Fig. 9 we see that the height of the peaks monotonously 
decreases and afterwards it begins to oscillate. We do
not have an explanation of this fact. For even longer times we see that
there does not exist a definite pattern repeating itself (Fig. 10, notice that in
this picture only the envelopement of the peaks is plotted). 
Fig. 11 shows the form of a peak ($N+1=32$) which is
non-symmetrical. The growing side of the peak is steeper than the
subsequent quasi-exponential decay. In Fig. 12 we choose $D=20A.$ In
this case conditions (\ref{cp}) are not satisfied and a non-dissipative
behavior occurs with very quick oscillations. For Figs. 13 and 14 we have
selected the value $D=2A$ for $N+1=32$ and $100$ respectively. This value
satisfies conditions (\ref{cp}) [remember that the bound (\ref{cota}) is
excessive]. However this worsening of $D$ is reflected in the fact that $%
\left\langle N_\Omega \right\rangle $ presents more fluctuations.

\vspace{0.5in}
\epsfysize=7.5truecm
\centerline{\epsffile{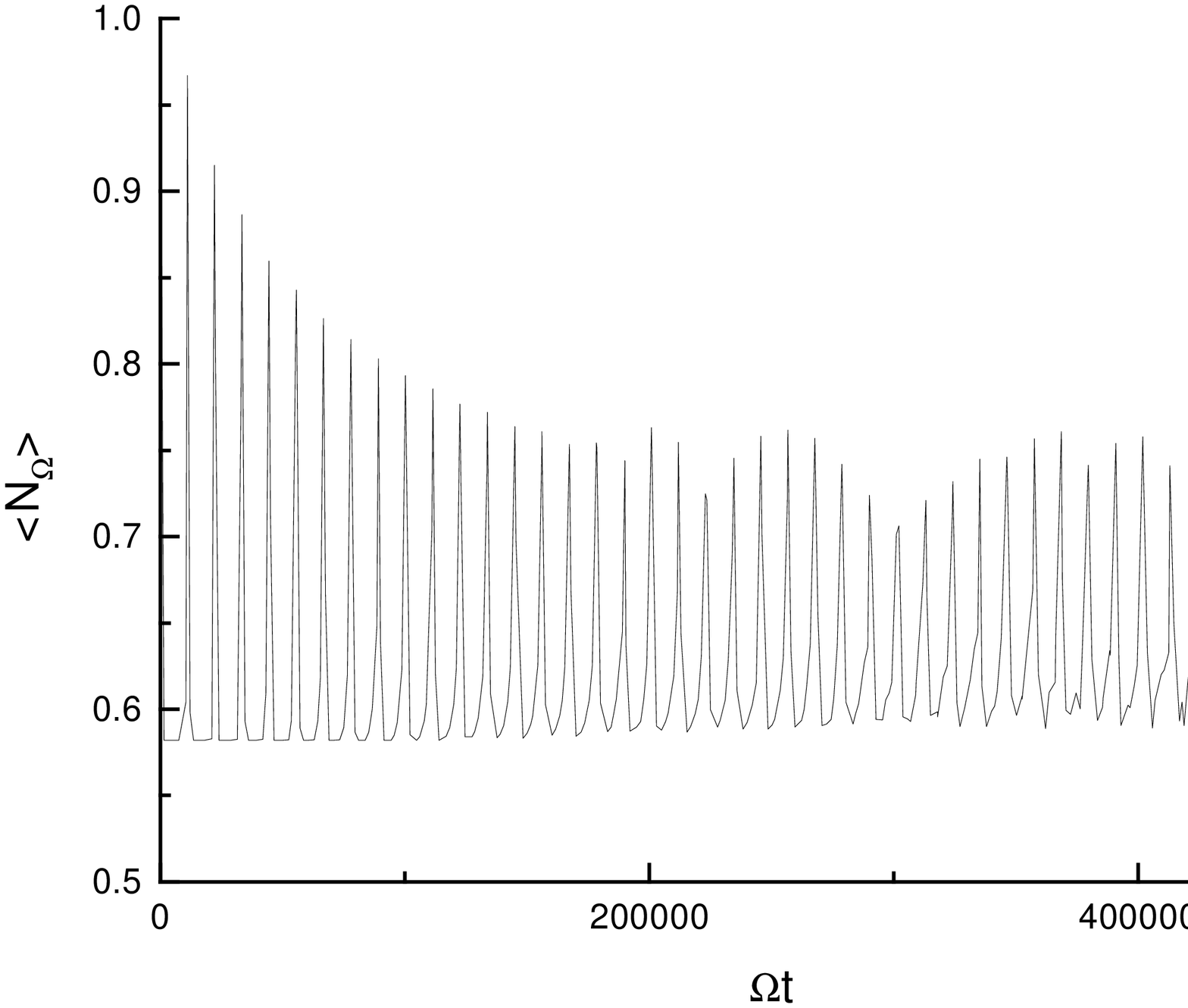}\hspace{1.15in}\epsfysize=7.5truecm\epsffile{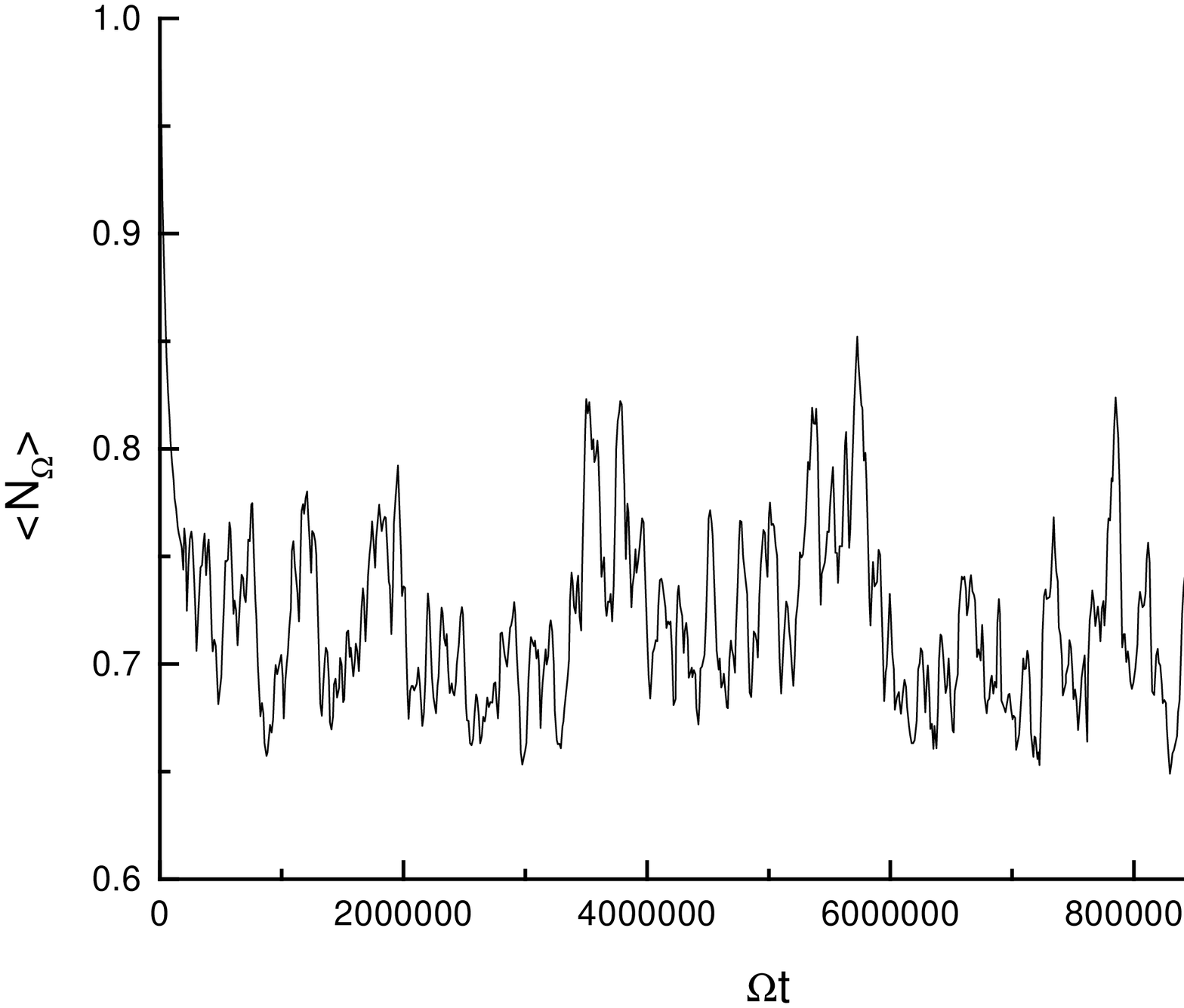}}
\centerline{\ \ \ \ \ \ {\bf FIG.    9.}\  Behvior  of the peaks of $\left\langle N_\Omega
\right\rangle$. \hspace{1.in} {\bf FIG. 10.}\ Peaks for long times.}
\vspace{0.6in}
\epsfysize=7.5truecm
\centerline{\epsffile{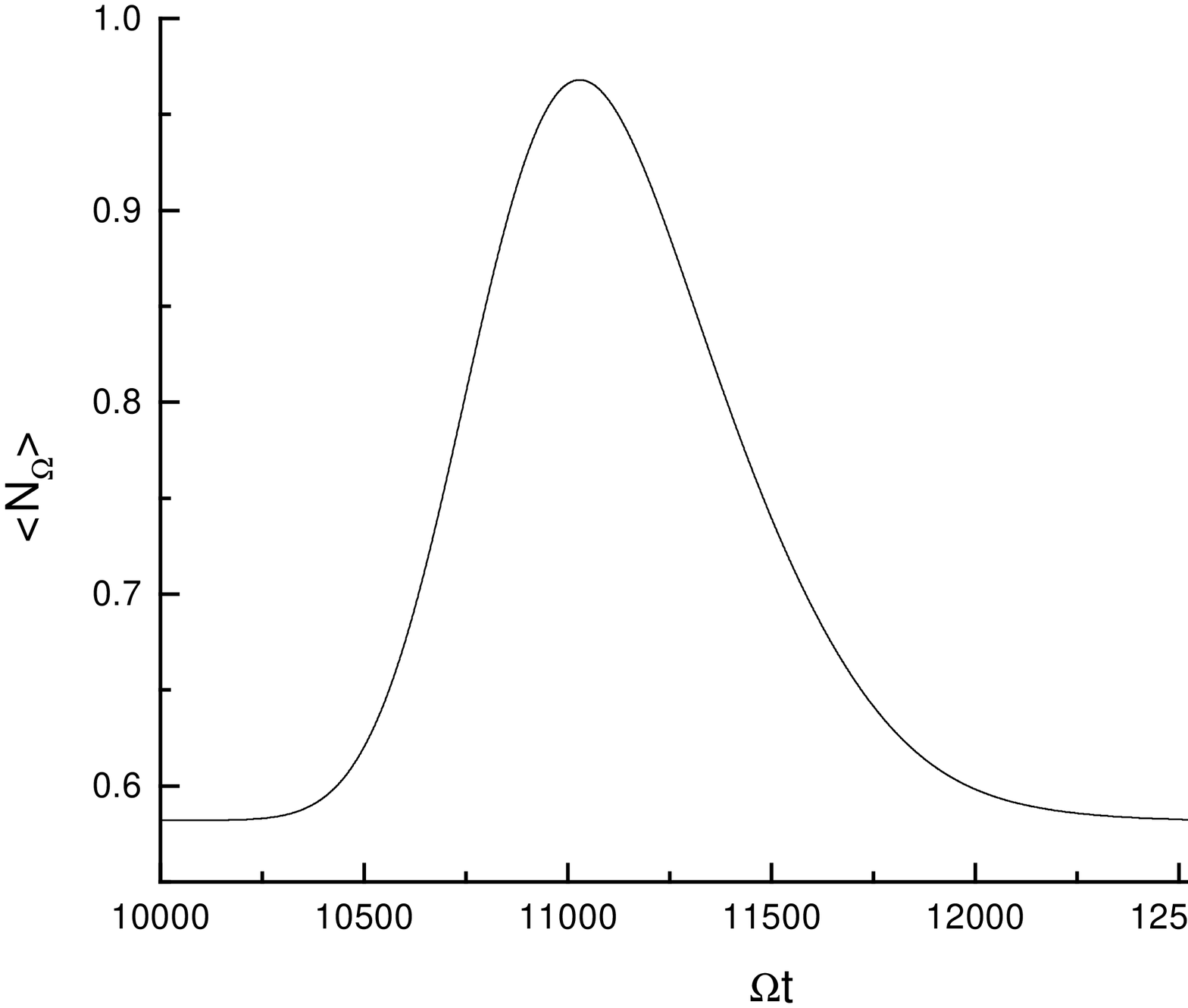}\hspace{1.15in}\epsfysize=7.5truecm\epsffile{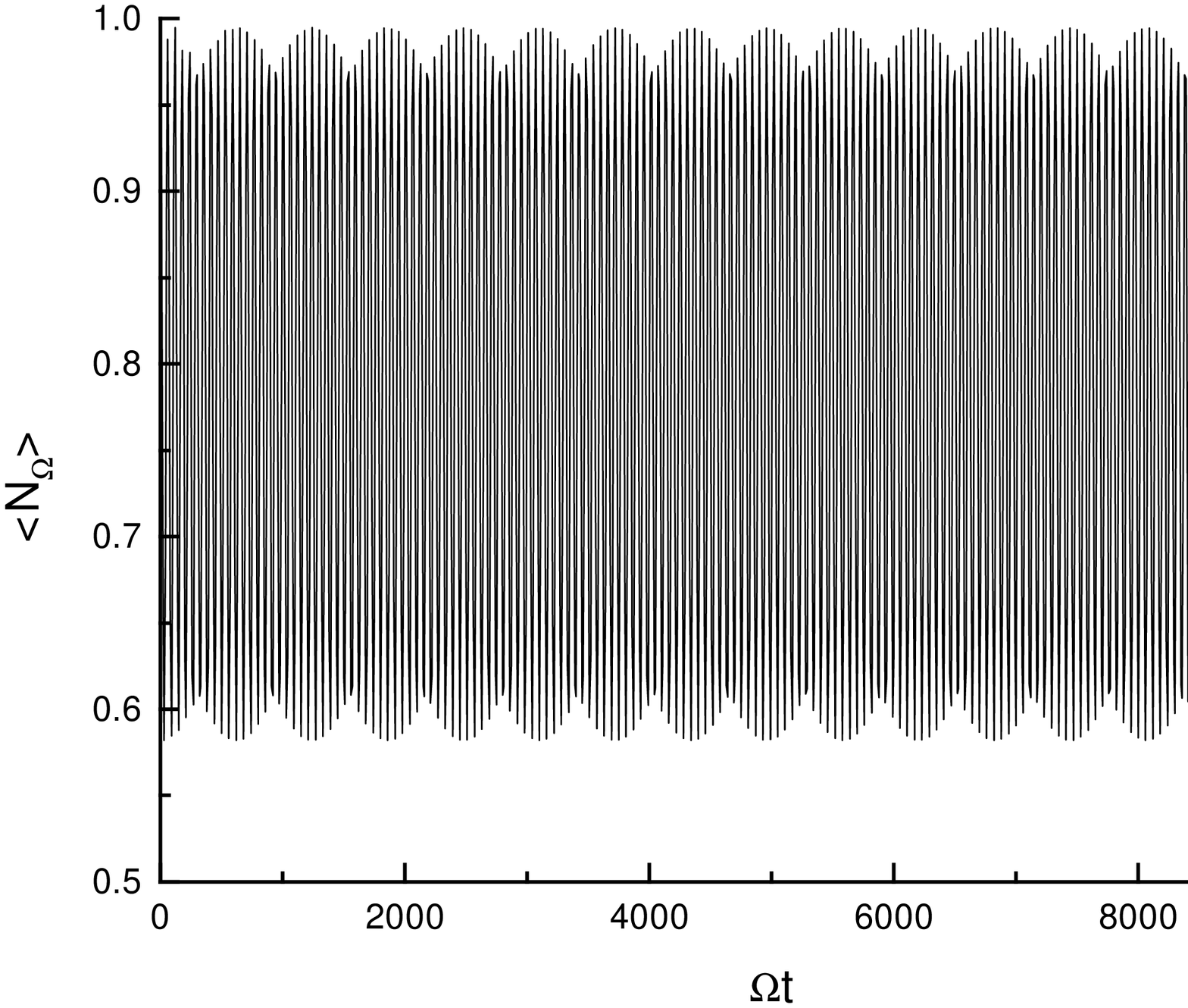}}
\centerline{\ \ \ {\bf FIG.   11.}\  Form  of  a  peak.   \hspace{1.5in} {\bf FIG.
12.}\ $N+1=32$, $D=20A$.}
\vspace{0.6in}
\epsfysize=7.5truecm
\centerline{\epsffile{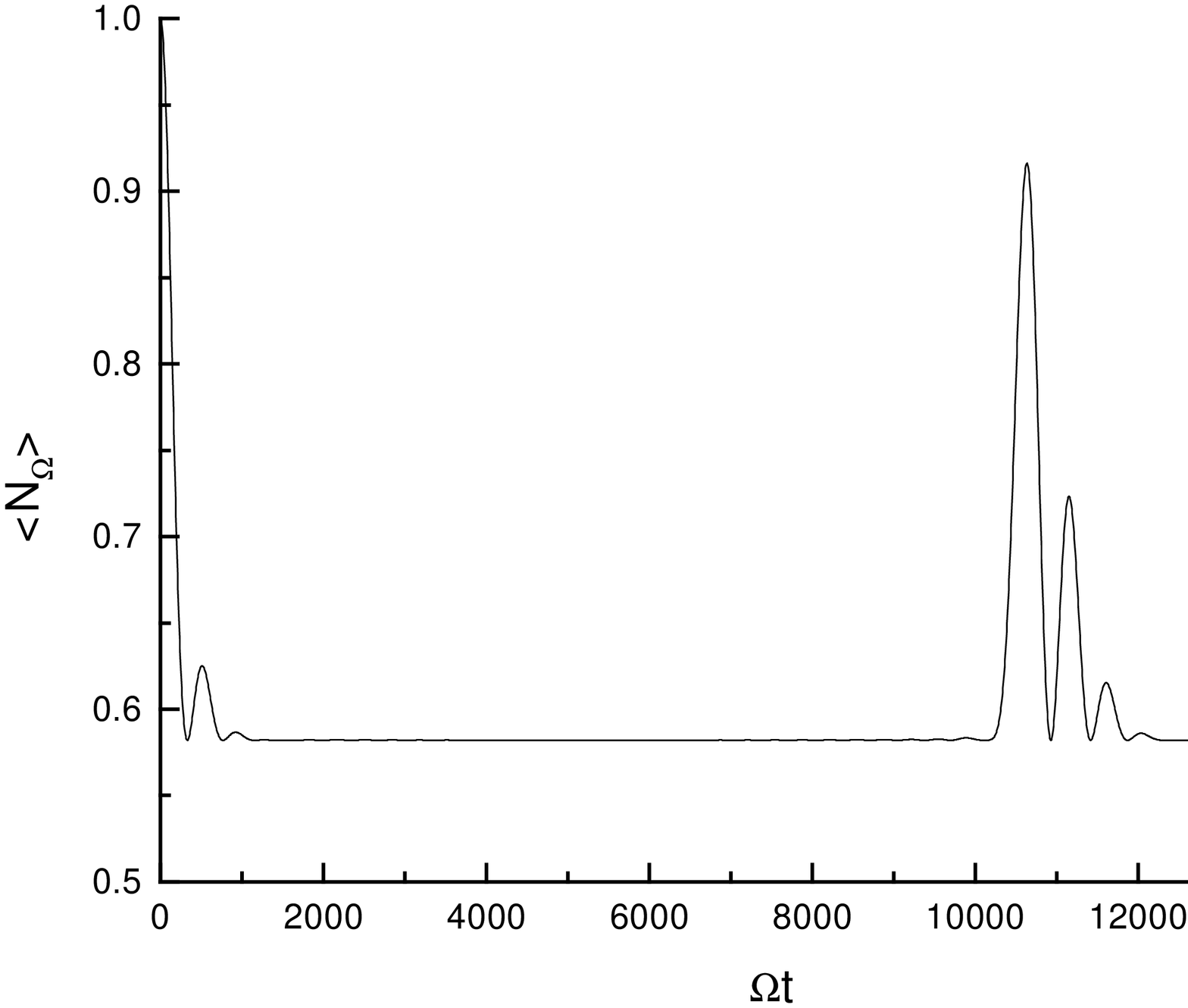}\hspace{1.15in}\epsfysize=7.5truecm\epsffile{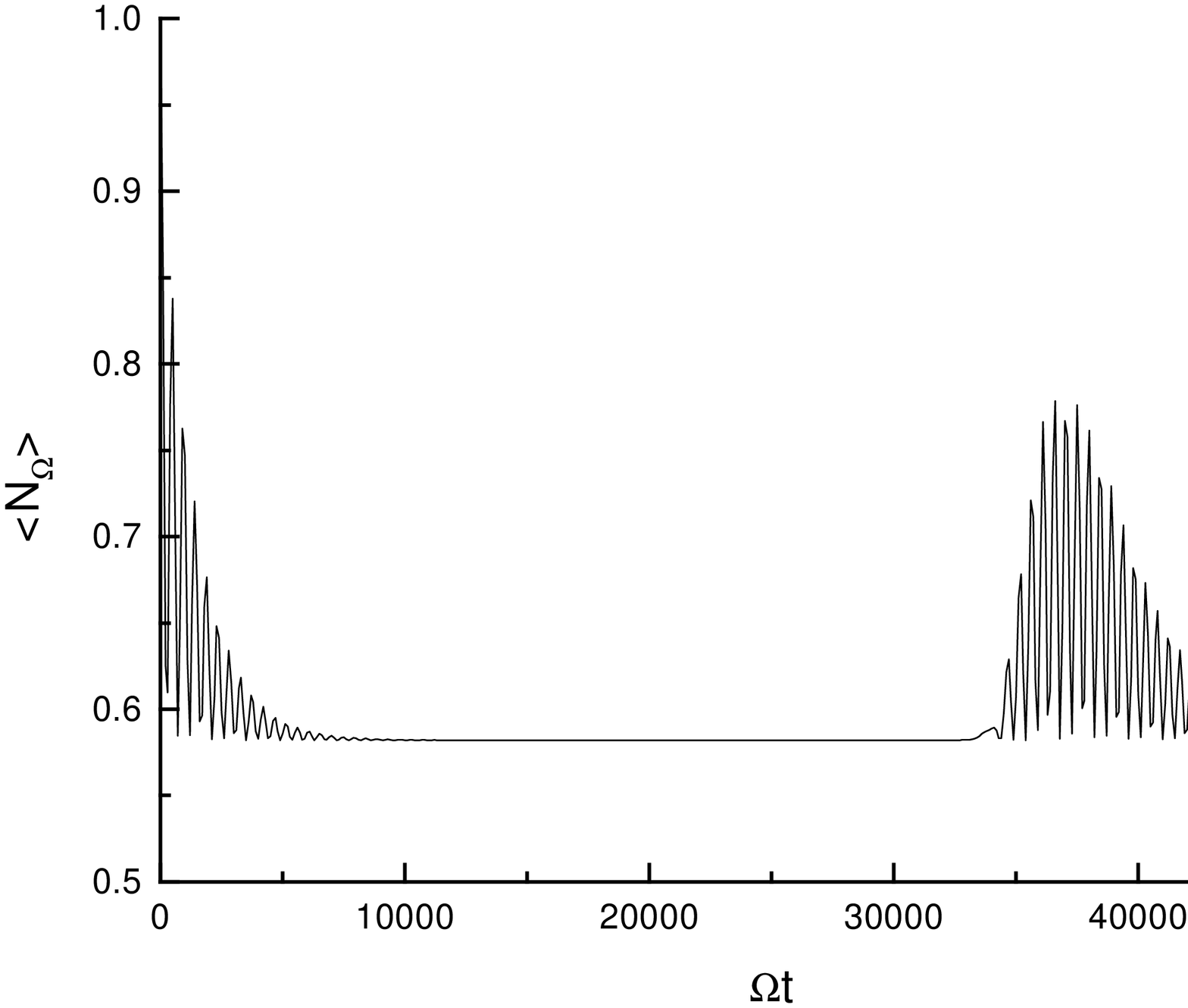}}
\centerline{\ \ \ \ \ \ {\bf FIG.  13.}\  $N+1=32$,  $D=2A$.    \hspace{1.8in}  {\bf  FIG.
14.}\ $N+1=100$, $D=2A$.}
\vspace{0.6in}

In Figs. 15 to 18 we plot $\left\langle X\right\rangle $ superposed to $%
\left\langle N_\Omega \right\rangle $ for times around the second
peak centered in $t_P.$ We see how the behavior of $\left\langle
X\right\rangle $ and $\left\langle N_\Omega \right\rangle $ are correlated
and how after the revival the subsystem oscillator is damped [see Eq. (74)
in Sec. VI]. In Fig. 16 the choice of $D=2A$ shows again the growth of
the fluctuations.

\vspace{0.5in}
\epsfysize=7.5truecm
\centerline{\epsffile{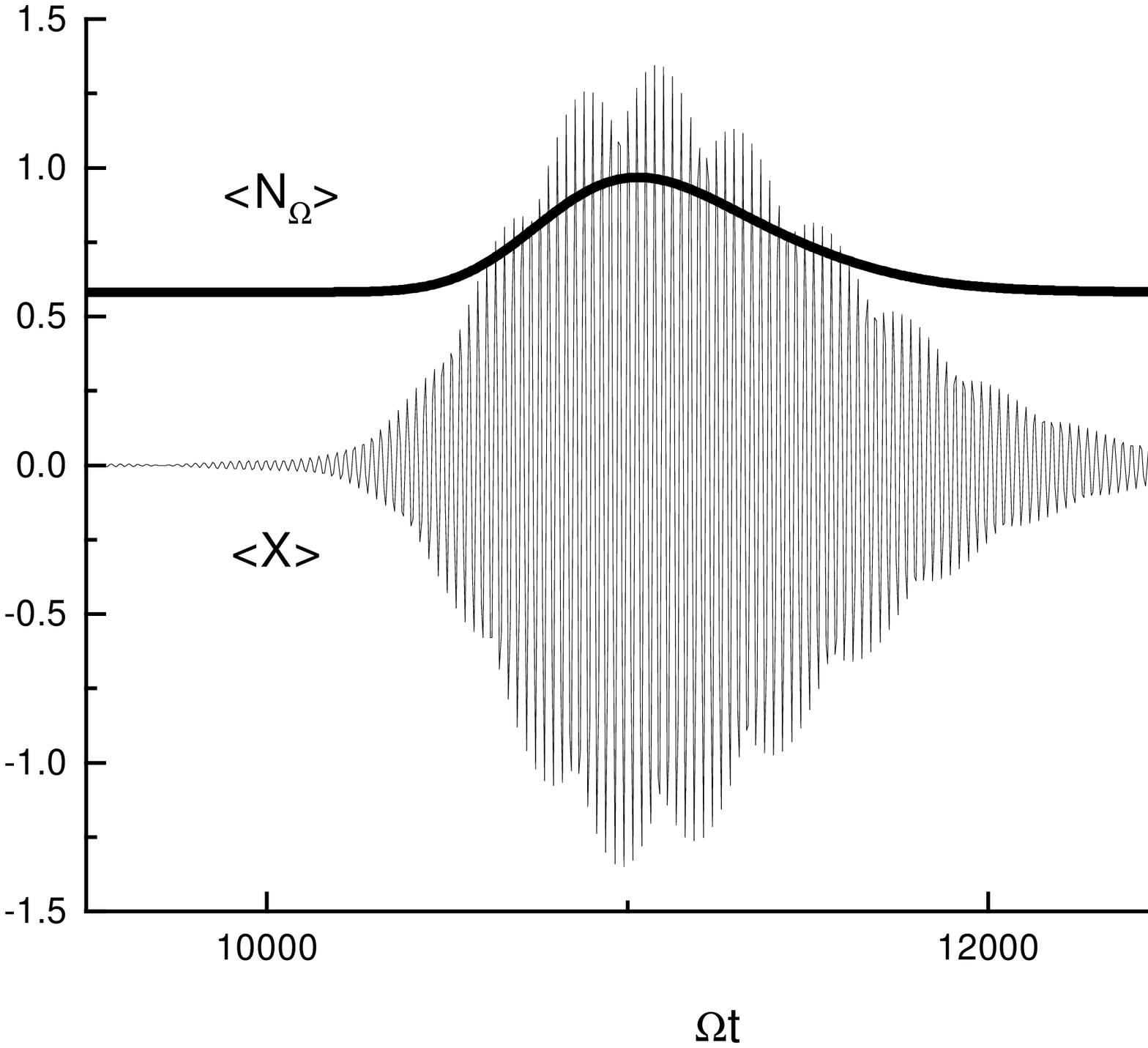}\hspace{1.15in}\epsfysize=7.5truecm\epsffile{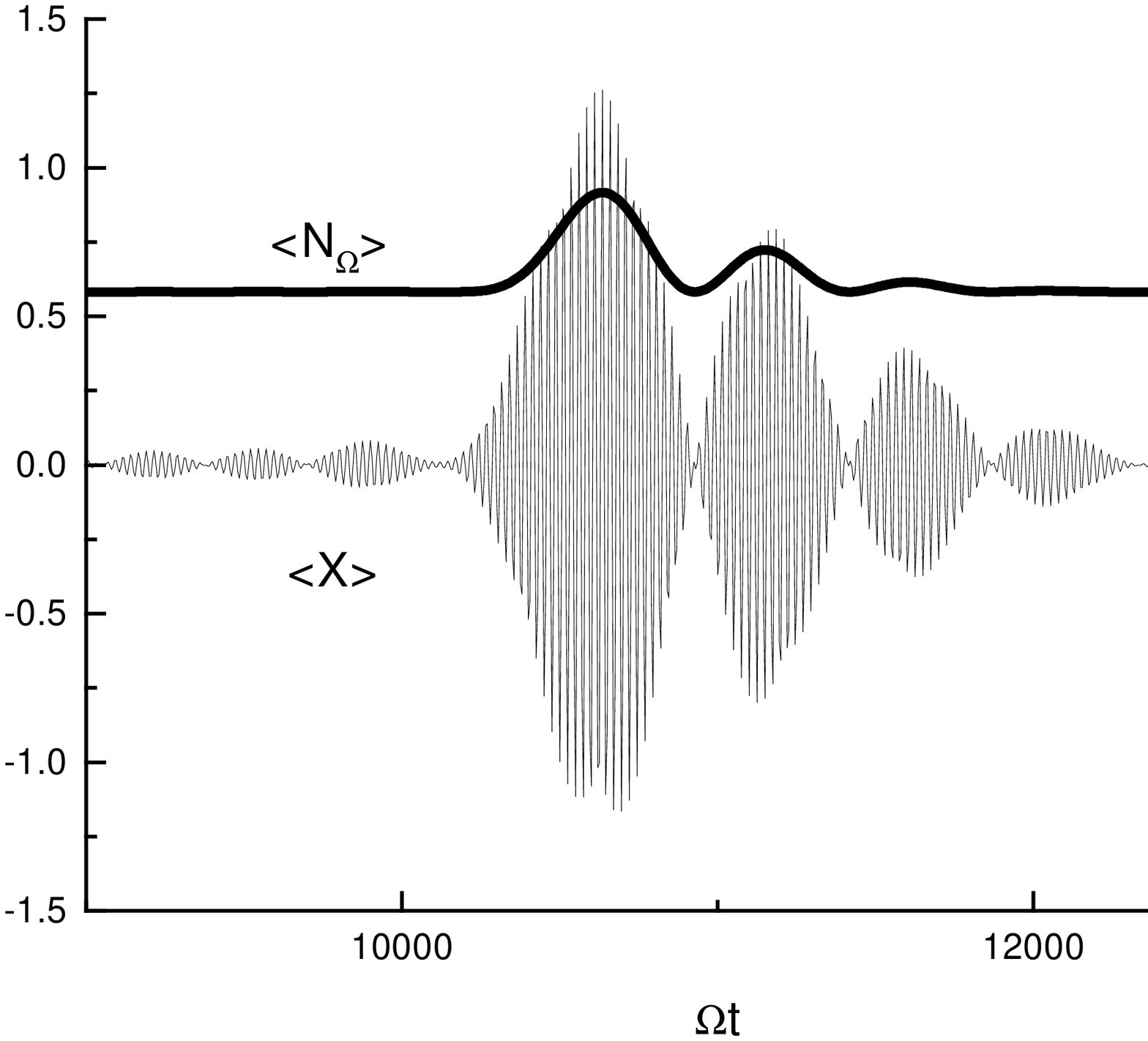}}
\centerline{\ \ \ \ \ \ {\bf FIG.  15.}\  $N+1=32$, $D=A$.    \hspace{2.in}  {\bf  FIG.
16.}\ $N+1=32$, $D=2A$.}
\vspace{0.6in}
\epsfysize=7.5truecm
\centerline{\epsffile{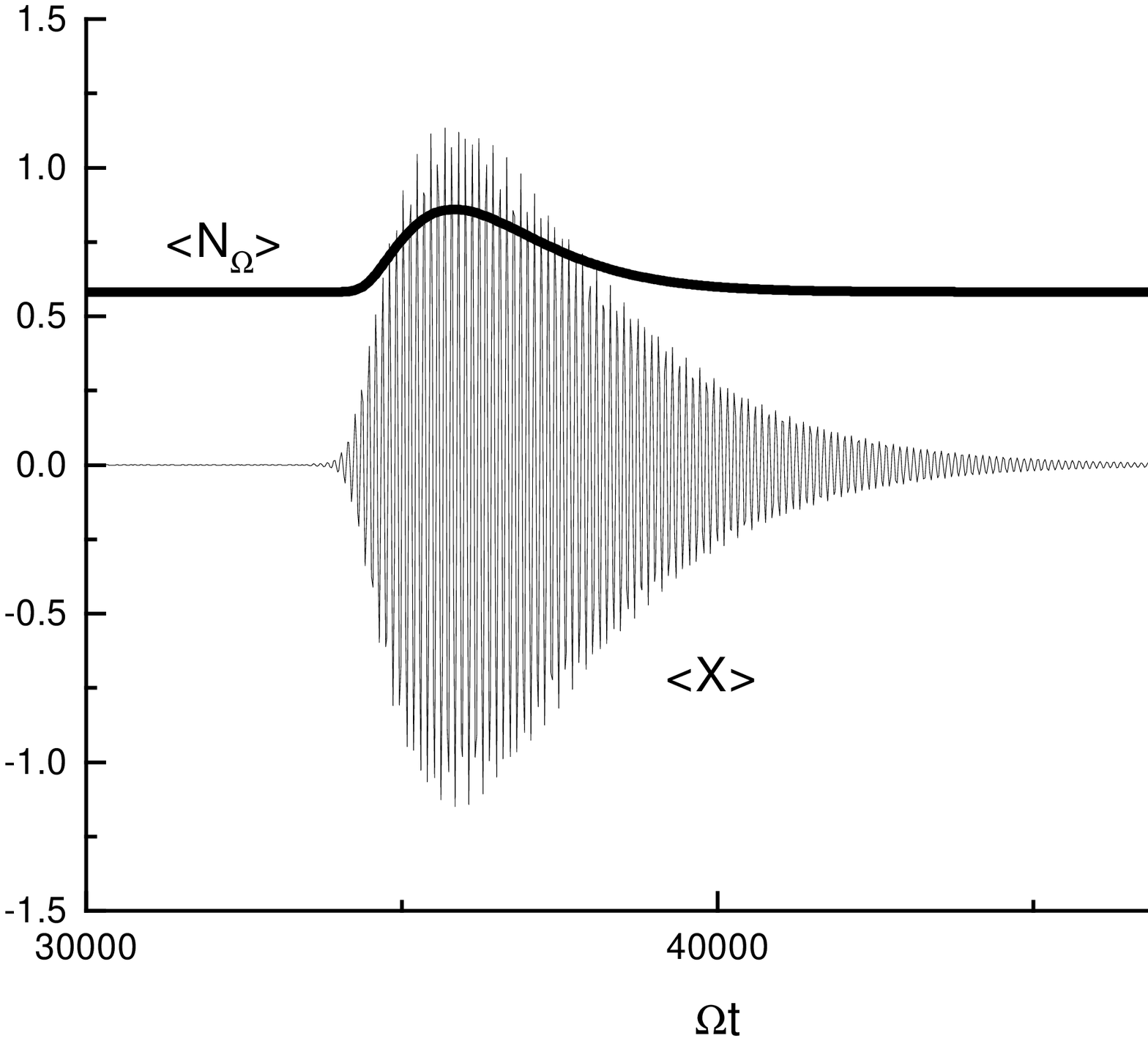}\hspace{1.15in}\epsfysize=7.5truecm\epsffile{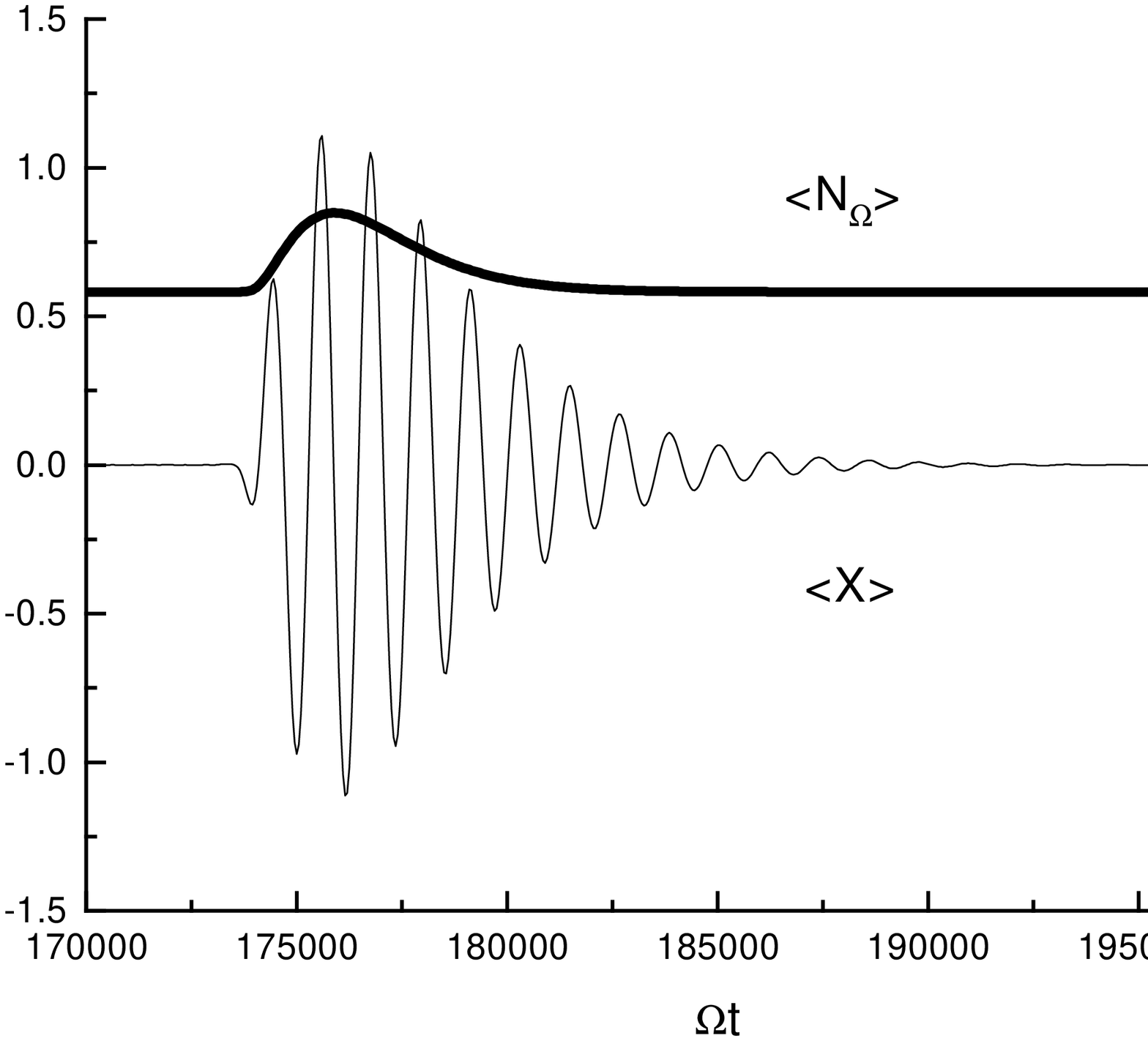}}
\centerline{\ \ \ \ \ {\bf FIG.  17.}\  $N+1=100$.    \hspace{1.8in}  {\bf  FIG.
18.}\ $N+1=500$.}
\vspace{0.6in}

Fig. 19 contains the behavior of $P_{\Omega \Omega }$ and $P_{\Omega n}$ for 
$N+1=32.$ We see that the bath contribution remains almost constant except at
times around $t_P$ in which the survival probability has a maximum.
In Fig. 20 we compare the behavior of $\left\langle N_n\right\rangle $ for a
value near to the central frequency $\Omega $ and for a value far of it. We
see that while $\left\langle N_2\right\rangle $ smoothly fluctuates around a
constant value $\left( e^{\beta \omega _2}-1\right) ^{-1}$, $\left\langle
N_{16}\right\rangle $ is sensible to what happens with $\left\langle
N_\Omega \right\rangle $ and then it is displaced with respect to $\left( e^{\beta
\omega _{16}}-1\right) ^{-1}.$ This is an indication that the transference of
energy from the subsystem to the bath is more effective for frequencies
near to $\Omega$. 
In Fig. 21 we confirm the hypothesis that the central
oscillators are those which receive the energy of the Brownian particle,
since going to the continuous limit the distribution of $P_{\Omega n}$
approaches to a delta function. The asymptotic value of $\left\langle
N_\Omega \right\rangle $ is given by

\begin{equation}
\left\langle N_\Omega (\infty )\right\rangle =\sum\limits_{n=1}^NP_{\Omega
n}(\infty )\frac 1{e^{\beta \omega _n}-1},\label{asidis}
\end{equation}
where $P_{\Omega n}(\infty )=\sum\limits_{\nu =0}^N\left( \left| \Phi _\nu
\right| ^2\frac{g_n}{\alpha _\nu -\omega _n}\right) ^2\equiv\theta _N\left(
\omega _n\right) .$ In the continuum we show, in Sec. VI, that $P_{\Omega
\omega }(\infty )=\delta (\omega -\Omega )$ up to the first order in a re-scaled  
parameter $\tau=\lambda^2 t$ \cite{van Hove} (which is known as $\lambda^2 t$ approximation). 
Fig. 21 thus plots $\theta _N$ ${\it vs.}$ $%
\omega _n.$

\vspace{0.5in}
\epsfysize=7.5truecm
\centerline{\epsffile{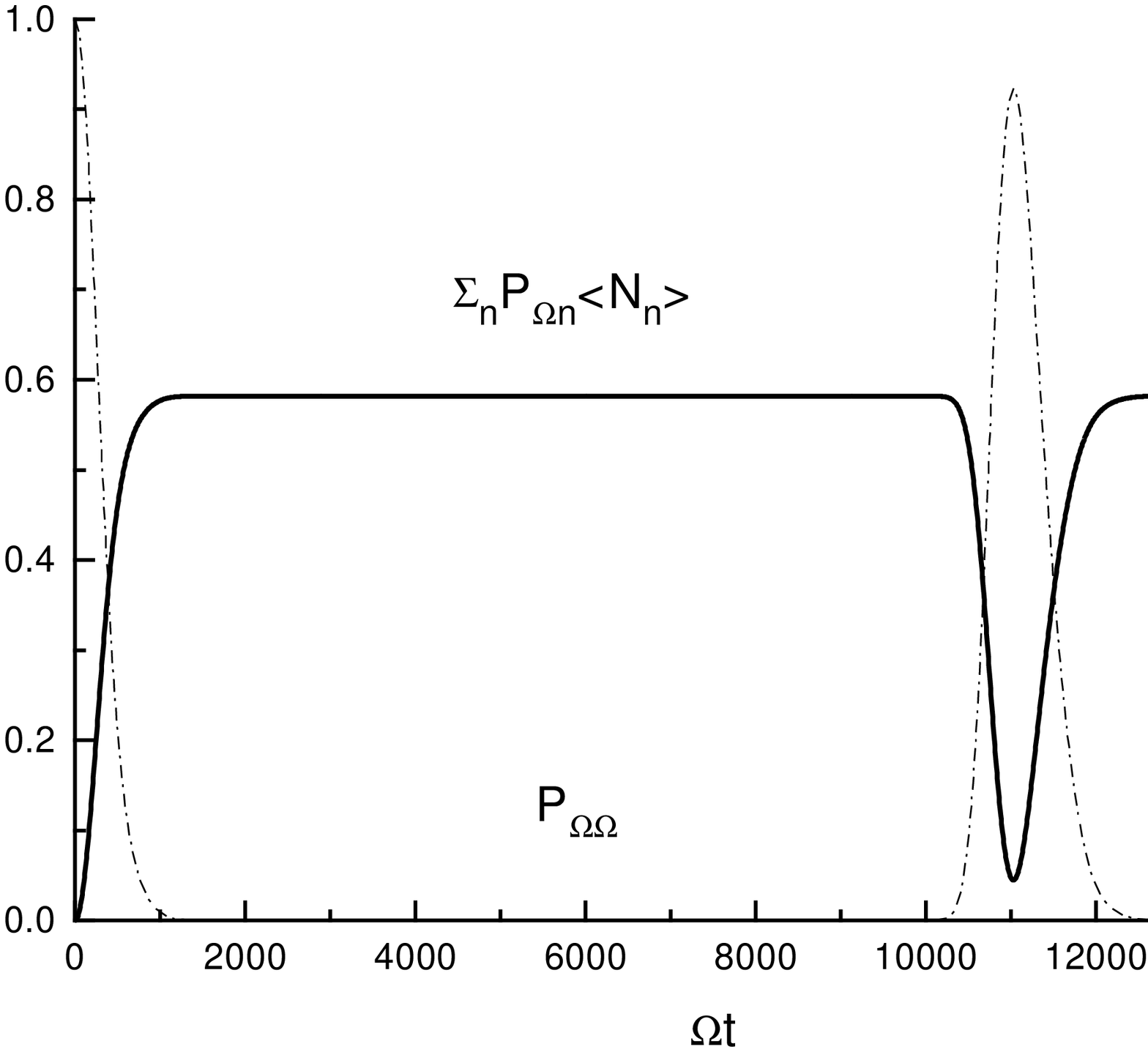}\hspace{1.15in}\epsfysize=7.5truecm\epsffile{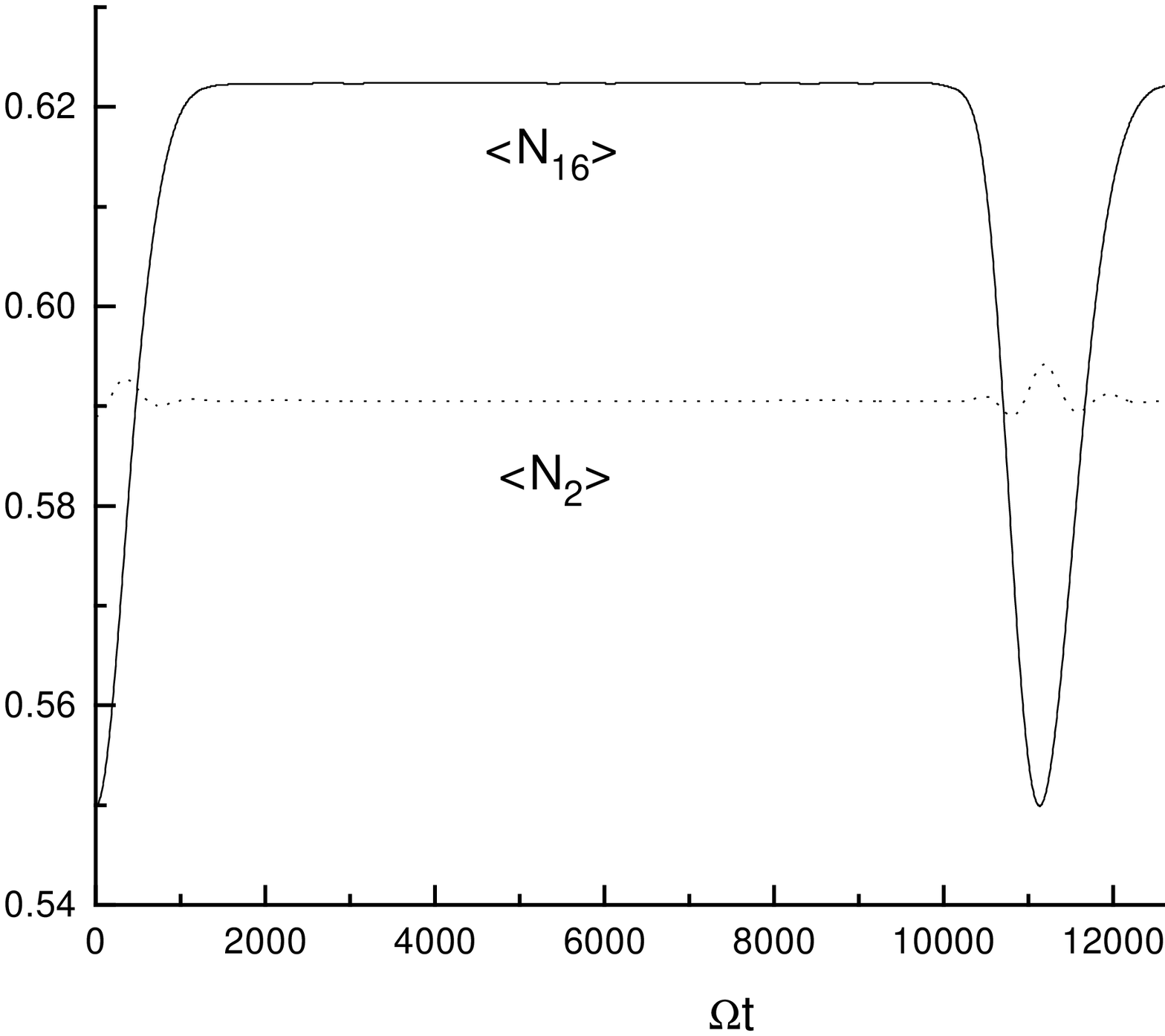}}
\centerline{{\bf  FIG.        19.}\   Survival  probability and  bath
contribution.  \hspace{0.6in}  {\bf  FIG.
20.}\ Bath population.}
\vspace{0.6in}
\epsfysize=12truecm
\centerline{\epsffile{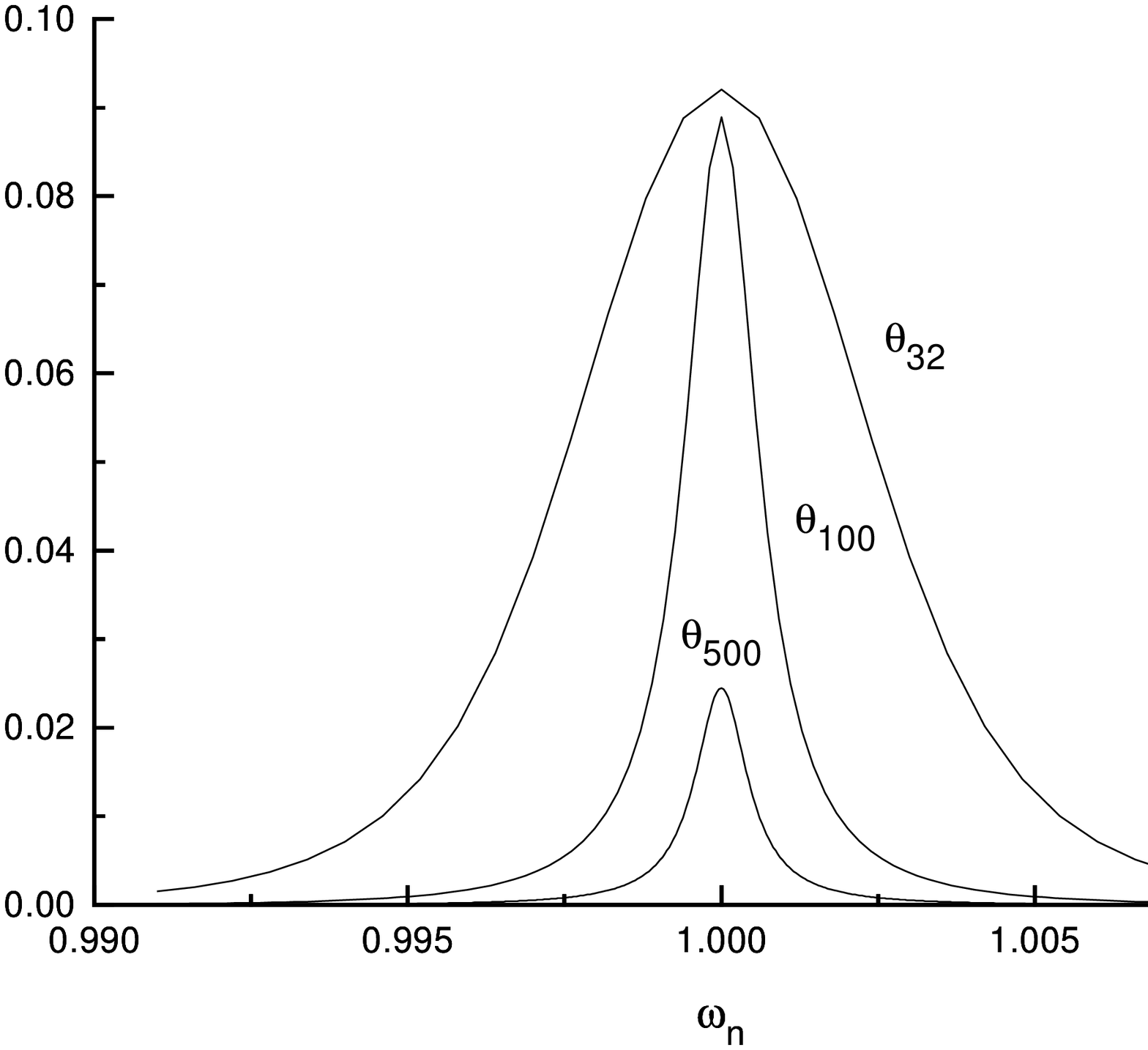}}
\centerline{\ \ \ {\bf FIG.  21.}\  Energy transfer to the central bath oscillator.}
\vspace{0.6in}

We end this section by obtaining the results of Ref. \cite{Gruver} for $N+1=32.$ In Fig. 22 we
show $\left\langle N_\Omega \right\rangle $ ${\it vs.}$ $\Omega t$ (cf. Fig.
1 (b) of Ref. \cite{Gruver}). In Fig. 23 we see the damping oscillations, in
Fig. 24 we compare the survival probability with the bath contribution, and
in Fig. 25 we show the behavior of different values of $\left\langle
N_n\right\rangle .$

\vspace{0.5in}
\epsfysize=7.5truecm
\centerline{\epsffile{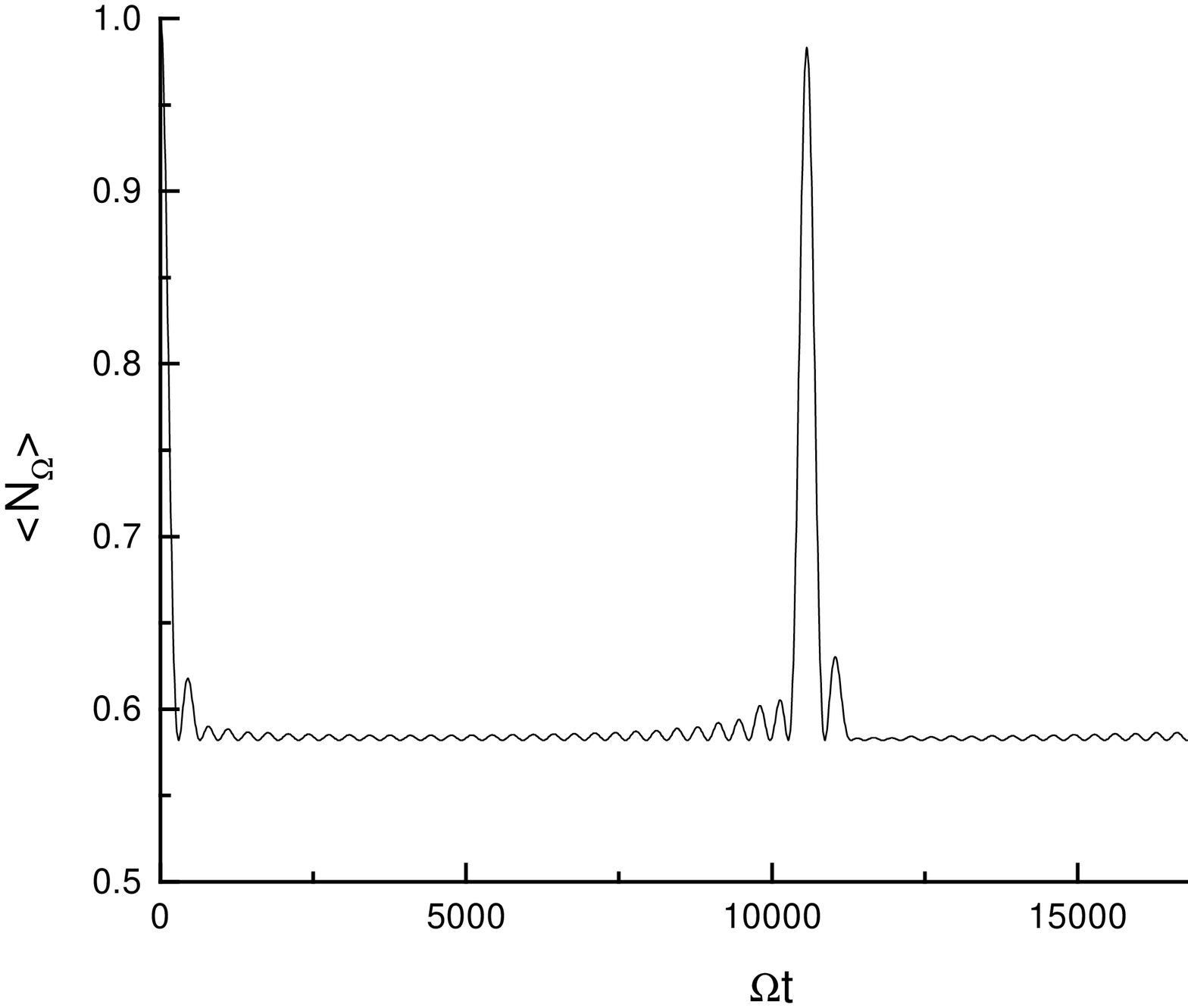}\hspace{1.15in}\epsfysize=7.5truecm\epsffile{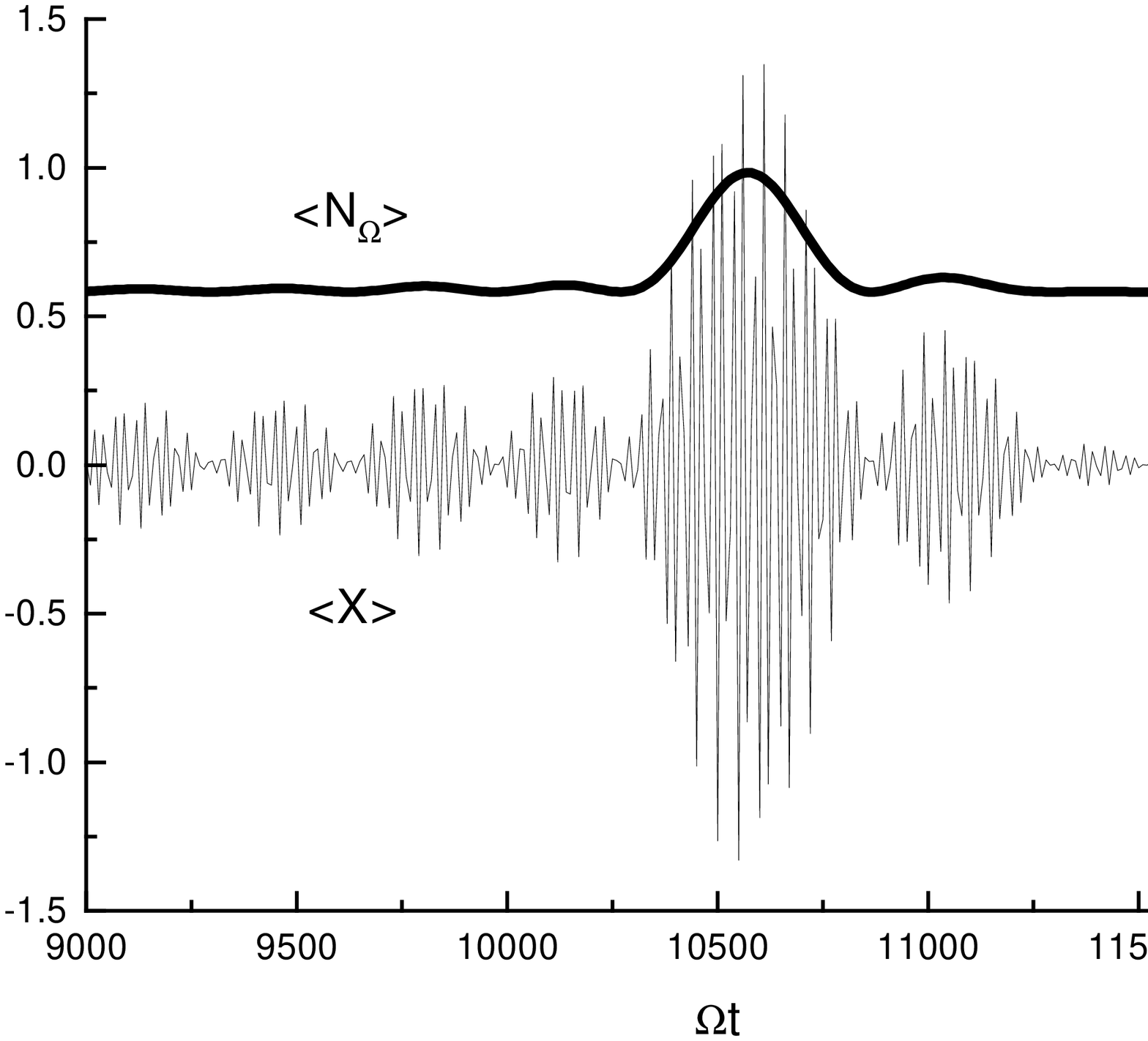}}
\centerline{\ \ \ \ {\bf FIG.   22.}\  $\left\langle N_\Omega \right\rangle$ {\it vs.}
$\Omega t$.    \hspace{1.7in}  {\bf  FIG.
23.}\ $\left\langle X \right\rangle$ {\it vs.} $\Omega t$.}
\vspace{0.6in}
\epsfysize=7.5truecm
\centerline{\epsffile{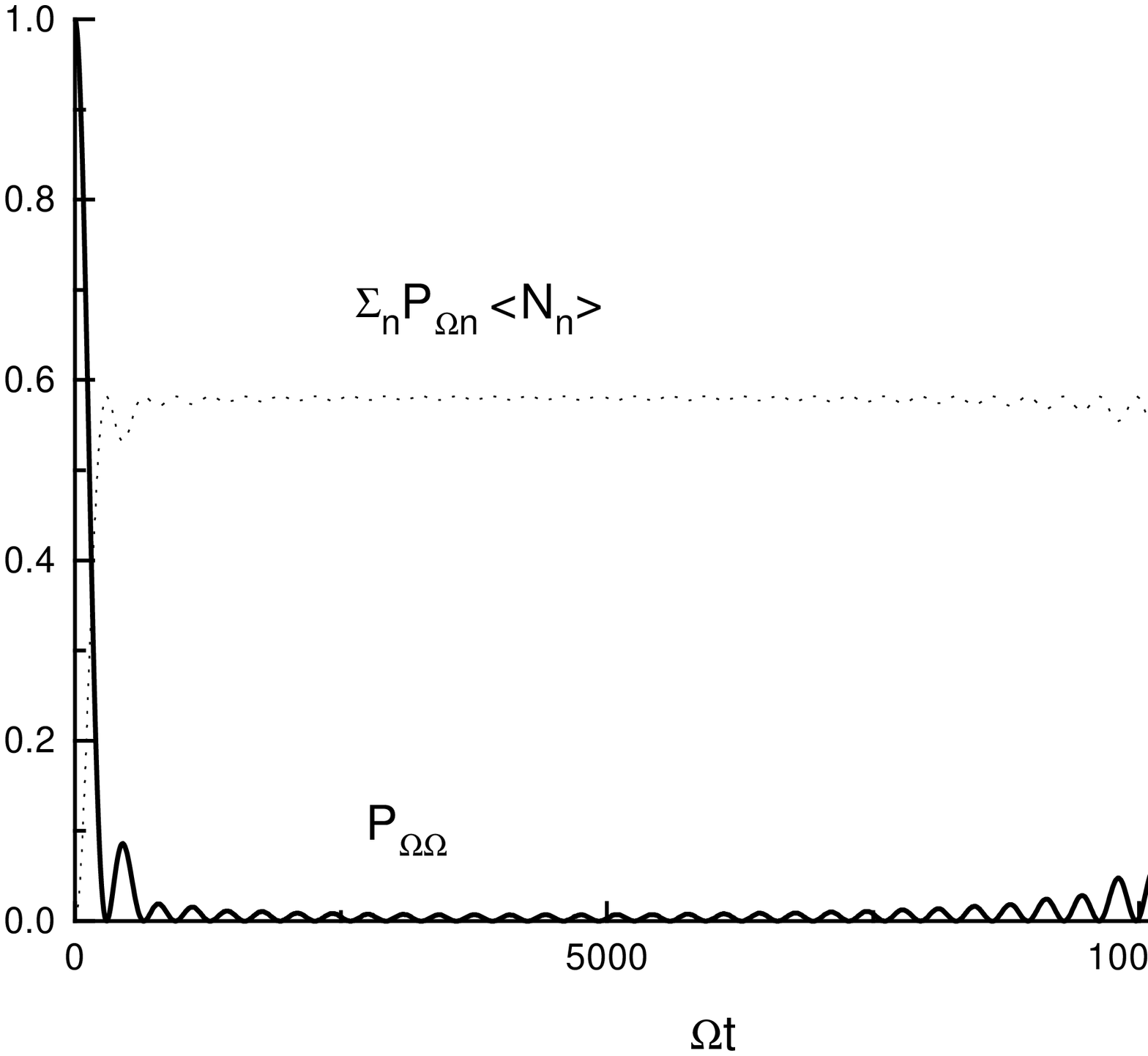}\hspace{1.15in}\epsfysize=7.5truecm\epsffile{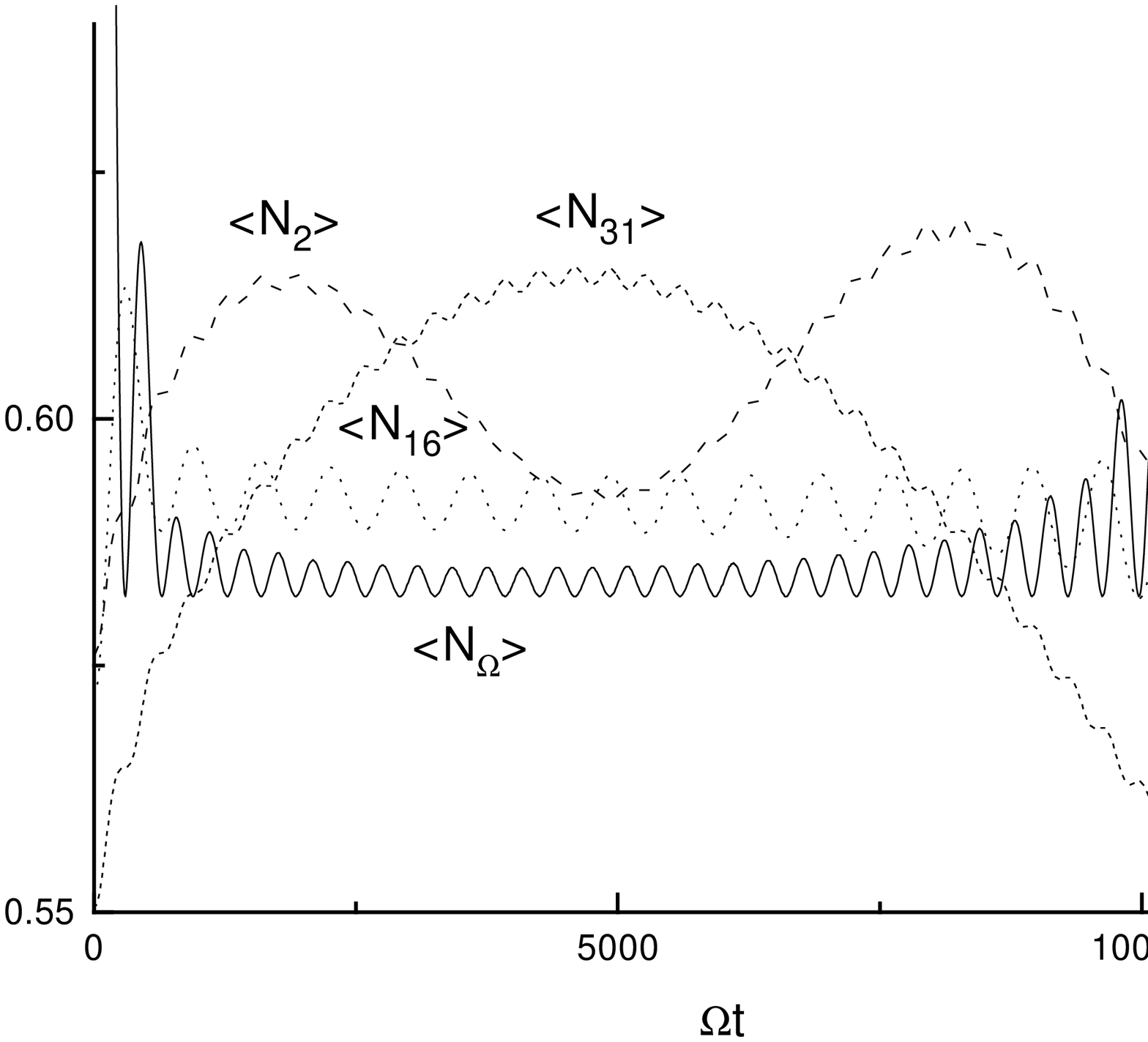}}
\centerline{{\bf FIG.  24.}\ Survival probability and bath contribution.    \hspace{0.3in}  {\bf  FIG.
25.}\ $\left\langle N_n \right\rangle$ {\it vs.} $\Omega t$.}
\vspace{0.6in}

\medskip
\section{Asymptotic limit and continuous bath}
\medskip

In Sec. V we have studied a particular model for a finite but increasing
number of bath oscillators with the aim of approaching to the
continuous limit. In this section we analytically take this limit for 
frequencies spanning a segment into the positive real axis. At first
glance the continuous limit of Eq. (\ref{sfci}) can be made by changing
summations for integrals. However it cannot be straightforwardly taken 
because of
the appearance of a continuous set of divergencies along the integration
domain. We must give a criterion to avoid these singularities, which is
related to the choice of the boundary conditions. Following Ullersma's
pioneering work \cite{Ullersma} we propose a way, based on an analytic
continuation method, to do that.

We can define the function $R_d(z)$ (the reduced resolvent operator in the
energy representation, where $d$ stands for discrete case) of the complex
variable $z$ departing from Eq. (\ref{ea}) as

\begin{equation}
R_d^{-1}(z)=z-\Omega -\sum\limits_{n=1}^N\frac{g_n^2}{z-\omega _n},
\label{rz}
\end{equation}
where the normal frequencies $\alpha _\nu $ are given by the simple poles of 
$R_d:$ $R_d^{-1}(\alpha _\nu )=0.$ Eq. (\ref{nor}) can be rewritten in terms
of $R_d$ as

\begin{equation}
\left| \Phi _\nu \right| ^2=\frac 1{\left( R_d^{-1}\right) ^{\prime }(\alpha
_\nu )}.  \label{norr}
\end{equation}
Eq. (\ref{norr}) allows us to write the first equation of Eq. (\ref{atbtm})
as

\begin{equation}
B(t)=\sum\limits_{\nu =0}^N\frac{e^{-i\alpha _\nu t}}{\left( R_d^{-1}\right)
^{\prime }(\alpha _\nu )}\left[ B(0)+\sum\limits_{n=1}^N\frac{g_n}{\alpha
_\nu -\omega _n}b_n(0)\right] .  \label{yava}
\end{equation}
In order to perform the continuous limit let us consider the following
identities:

\begin{equation}
\sum\limits_{\nu =0}^N\frac{e^{-i\alpha _\nu t}}{\left( R_d^{-1}\right)
^{\prime }(\alpha _\nu )}=\frac 1{2\pi i}\oint_C dz\frac{e^{-izt}}{R_d^{-1}(z)},
\label{lapri}
\end{equation}

\begin{equation}
\sum\limits_{\nu =0}^N\frac{e^{-i\alpha _\nu t}}{\left( R_d^{-1}\right)
^{\prime }(\alpha _\nu )}\sum\limits_{n=1}^N\frac{g_n}{\alpha _\nu -\omega _n%
}b_n(0)=\frac 1{2\pi i}\oint_C dz\frac{e^{-izt}}{R_d^{-1}(z)}\sum\limits_{n=1}^N%
\frac{g_n}{z-\omega _n}b_n(0),  \label{lase}
\end{equation}
where $C$ is a counterclockwise contour in the $z$-plane that encircles the $%
N+1$ singularities of $R_d$ in the positive real axis (see Fig. 26).

\epsfysize=10truecm
\centerline{\epsffile{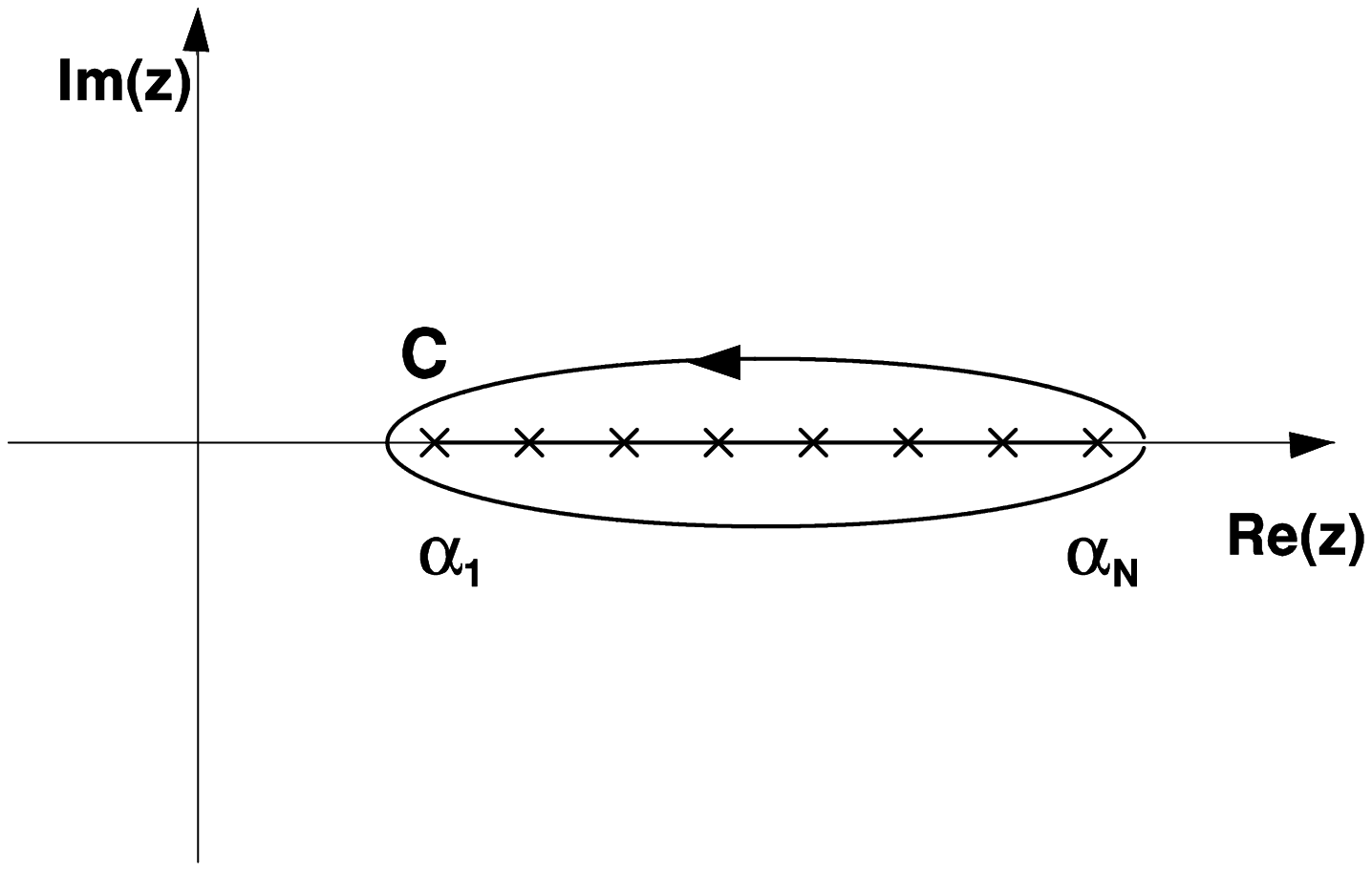}}
\centerline{{\bf FIG. 26.}\ Counterclockwise contour $C$.}
\bigskip
\bigskip

We make use of the residues theorem: $\oint_C\frac{P(z)}{Q(z)}dz=2\pi
i\sum_{k=1}^r{\rm Res}[\frac PQ,z_k]=2\pi i\sum_{k=1}^r\frac{P(z_k)}{%
Q^{\prime }(z_k)},$ where $z_k$ are the simple zeros of $Q(z).$ Eq. (\ref
{lapri}) is a direct consequence of the residues theorem and Eq. (\ref{lase}%
) follows from the same theorem and from the fact that $\sum_n\frac{%
e^{-i\omega _nt}}{R_d^{-1}(\omega _n)}g_nb_n(0)=0,$ since $R_d^{-1}(\omega
_n)$ diverges for all $n$.

When the bath frequencies form a dense set, the normal frequencies are also
dense. This limit of a continuous bath is valid for times that satisfy 

\[
t\ll {\rm \min }\left( \alpha _{\nu +1}-\alpha _\nu \right) ^{-1}. 
\]
In this approximation $R_d^{-1}(z)$ goes to

\begin{equation}
R^{-1}(z)=z-\Omega -\int_{\omega _{\min }}^{\omega _{\max }}d\omega \frac{%
g^2(\omega )}{z-\omega },  \label{alfaz}
\end{equation}
where $\omega _{\min }$ and $\omega _{\max }$ are the extrema of the
continuous set (lower and upper cutoff respectively) and avoid infrared and
ultraviolet divergencies respectively. \\

%\newpage
$g^2(\omega )$ is defined by \cite
{Ullersma}

\[
g^2(\omega )\Delta \omega =\sum\limits_{\omega <\omega _n<\omega +\Delta
\omega }g_n^2.
\]
The function $R^{-1}(z)$ has a cut along $\left( \omega _{\min },\omega
_{\max }\right) $, corresponding to the continuous spectrum of normal
frequencies. In order to ensure that the equation $R^{-1} (z)=0$ has no real 
roots it is
necessary that $\Omega \in \left( \omega _{\min },\omega _{\max }\right) $,
which together with conditions (\ref{cp}) adequately generalized to this
case, provide a necessary and sufficient criterion of feasibility for
dissipation in linear models. Generalized conditions (\ref{cp}) are given by

\begin{equation}
\int_{\omega _{\min }}^{\omega _{\max }}d\omega \frac{g^2(\omega )}{\omega
-\omega _{\min } }<\Omega -\omega _{\min },\hspace{0.3in}%
\int_{\omega _{\min }}^{\omega _{\max }}d\omega \frac{g^2(\omega )}{\omega
_{\max }-\omega }<\omega _{\max }-\Omega ,  \label{cpc}
\end{equation}
assuming that the integrals are well defined. 
The cut of the function $R^{-1}(z)$ can be reached from above and below the
positive real axis, giving the limiting values $R^{-1}(\alpha \pm i\epsilon
)=\alpha -\Omega -\int_{\omega _{\min }}^{\omega _{\max }}d\omega \frac{%
g^2(\omega )}{\alpha -\omega \pm i\epsilon },$ where $\alpha \in \left(
\omega _{\min },\omega _{\max }\right) $. Then, contraction of the contour $%
C $ in Eq. (\ref{lapri}) yields

\begin{equation}
\frac 1{2\pi i}\oint_C\frac{e^{-izt}}{R^{-1}(z)}dz=\frac 1{2\pi i}\int_{\omega
_{\min }}^{\omega _{\max }}d\alpha e^{-i\alpha t}\left[ \frac 1{%
R^{-1}(\alpha -i\epsilon )}-\frac 1{R^{-1}(\alpha +i\epsilon )}\right] .
\label{papre}
\end{equation}
On the other hand, by taking into account the well known identity between
distributions

\begin{equation}
\frac 1{x\pm i\epsilon }={\rm PV}\frac 1x\mp i\pi \delta (x),  \label{ident}
\end{equation}
where PV stands for the principal value and $\delta $ is the Dirac delta
distribution, we have

\begin{equation}
R^{-1}(\alpha +i\epsilon )-R^{-1}(\alpha -i\epsilon )=\int_{\omega _{\min
}}^{\omega _{\max }}d\omega \frac{g^2(\omega )}{\alpha -\omega -i\epsilon }%
-\int_{\omega _{\min }}^{\omega _{\max }}d\omega \frac{g^2(\omega )}{\alpha
-\omega +i\epsilon }=2i\pi g^2(\alpha ).  \label{sip}
\end{equation}
Then using (\ref{sip}) Eq. (\ref{papre}) is reduced to

\begin{equation}
\frac 1{2\pi i}\oint_C\frac{e^{-izt}}{R^{-1}(z)}dz=\int_{\omega _{\min
}}^{\omega _{\max }}d\alpha \frac{g^2(\alpha )}{\left| R^{-1}(\alpha
+i\epsilon )\right| ^2}e^{-i\alpha t},  \label{pria}
\end{equation}
where we consider that $R^{-1}(\alpha -i\epsilon )=R^{-1*}(\alpha
+i\epsilon ).$ Contracting the contour $C$ in Eq. (\ref{lase}) and taking
the continuous limit we have

\[
\frac 1{2\pi i}\oint_C dz\frac{e^{-izt}}{R^{-1}(z)}\int_{\omega _{\min
}}^{\omega _{\max }}d\omega \frac{g(\omega )}{z-\omega }b_\omega (0)=\frac 1{%
2\pi i}\int_{\omega _{\min }}^{\omega _{\max }}d\alpha e^{-i\alpha
t}\int_{\omega _{\min }}^{\omega _{\max }}d\omega g(\omega ) 
\]

\[
\times \left[ \frac 1{R^{-1}(\alpha -i\epsilon )\left( \alpha -\omega
-i\epsilon \right) }-\frac 1{R^{-1}(\alpha +i\epsilon )\left( \alpha -\omega
+i\epsilon \right) }\right] b_\omega (0), 
\]
and after performing a straightforward calculation we obtain

\[
\frac 1{2\pi i}\oint_C dz\frac{e^{-izt}}{R^{-1}(z)}\int_{\omega _{\min
}}^{\omega _{\max }}d\omega \frac{g(\omega )}{z-\omega }b_\omega
(0)=\int_{\omega _{\min }}^{\omega _{\max }}d\alpha \frac{g(\alpha )}{%
R^{-1}(\alpha -i\epsilon )}e^{-i\alpha t}b_\alpha (0)  
\]

\begin{equation}
+\int_{\omega _{\min }}^{\omega _{\max }}d\alpha \int_{\omega _{\min
}}^{\omega _{\max }}d\omega \frac{g^2(\alpha )g(\omega )}{\left|
R^{-1}(\alpha +i\epsilon )\right| ^2\left( \alpha -\omega +i\epsilon \right) 
}e^{-i\alpha t}b_\omega (0),  \label{sea}
\end{equation}
where we have used identity (\ref{ident}) for $x=\alpha -\omega $ to have the $%
\delta $-function expressed as

\[
\frac 1{\left( \alpha -\omega -i\epsilon \right) }-\frac 1{\left( \alpha
-\omega +i\epsilon \right) }=2\pi i\delta (\alpha -\omega ). 
\]
Then $B(t)$ can be written in the continuous limit as

\begin{equation}
B(t)=\int_{\omega _{\min }}^{\omega _{\max }}d\alpha e^{-i\alpha t}\left[
\left| \Phi _\alpha \right| ^2B(0)+\Phi _\alpha ^{*}\int_{\omega _{\min
}}^{\omega _{\max }}d\omega \phi _\alpha (\omega )b_\omega (0)\right] ,
\label{fina}
\end{equation}
where

\begin{equation}
\Phi _\alpha =\frac{g(\alpha )}{R^{-1}(\alpha +i\epsilon )},  \label{coeflc}
\end{equation}

\begin{equation}
\phi _\alpha (\omega )=\delta (\alpha -\omega )+\frac{g(\alpha )g(\omega )}{%
R^{-1}(\alpha +i\epsilon )\left( \alpha -\omega +i\epsilon \right) }.
\label{coefcl}
\end{equation}
Compare Eqs. (\ref{coeflc}) and (\ref{coefcl}) with those obtained in Ref. 
\cite{Sudarshan} [Eqs. (4.8a) and (4.8b); in this work $\omega _{\min }=0$
and $\omega _{\max }=\infty $], which correspond to the Lippmann-Schwinger
coefficients of the eigenvectors of the continuous generalization of the
one-particle Hamiltonian (\ref{fri}) \cite{Gordo,Laura}.

We have obtained the continuous generalization of $B(t).$ 
By an straightforward but similar calculation we can obtain the continuous version 
of $b_n (t)$. 

From Eq. (\ref
{fina}) we can give an approximate expression of the Langevin equation. By
taking mean values in a thermal initial state for the bath we have

\begin{equation}
\left\langle B(t)\right\rangle =\int_{\omega _{\min }}^{\omega _{\max
}}d\alpha \left| \Phi _\alpha \right| ^2e^{-i\alpha t}\left\langle
B(0)\right\rangle .  \label{ame}
\end{equation}
Performing an analytical continuation to the complex plane, one can extract
the contribution of the poles of $\left| \Phi _\alpha \right| ^2.$ Then, let
us analyze the analytic structure of $\left| \Phi _\alpha \right| ^2$ as a
complex function. The function $R^{-1}(z)$ has no zeros in the complex
plane. It can be easily seen by considering $z=a+ib,$

\begin{eqnarray*}
a+ib-\Omega -\int_{\omega _{\min }}^{\omega _{\max }}d\omega \frac{%
g^2(\omega )}{a+ib-\omega } &=&a-\Omega -\int_{\omega _{\min }}^{\omega
_{\max }}d\omega \frac{(a-\omega )g^2(\omega )}{(a-\omega )^2+b^2} \\
&&+ib\left( 1+\int_{\omega _{\min }}^{\omega _{\max }}d\omega \frac{%
g^2(\omega )}{(a-\omega )^2+b^2}\right) ,
\end{eqnarray*}
where the imaginary part is equal to zero only if $b=0.$ Thus only real
zeroes can exist, but in this case $R^{-1}(z)$ is only well defined by its
limiting values $R^{-1}(\alpha \pm i\epsilon ).$ The discontinuity of $%
R^{-1}(z)$ in $\left( \omega _{\min },\omega _{\max }\right) $ is given by
Eq. (\ref{sip}). $R^{-1}(\alpha \pm i\epsilon )$ has no zeroes because the
imaginary part is not null, since $g^2(\alpha )\neq 0$ for $\alpha \in
\left( \omega _{\min },\omega _{\max }\right) .$ Then $R^{-1}(z),$ which is
analytic in the complex plane except for the cut discontinuity along $\left(
\omega _{\min },\omega _{\max }\right) ,$ has no zeroes in its definition
range. In order to define an analytic function in all the complex plane we
must analytically extend $R^{-1}(z)$ into the second Riemann sheet. To see
this let us consider the function $f(z)=(z-\omega )^{-1}$ with its limiting
values $(\alpha -\omega \pm i\epsilon )^{-1}.$ We can consider $f(z)$ as a
multivalued function or take the limiting values as defining two different
functions (complex distributions). Then we define

\begin{equation}
f_{\pm }(z)=\left\{ 
\begin{array}{l}
\frac 1{z-\omega },\hspace{1.3in}\ {\rm for}\ {\rm Im}z%
%TCIMACRO{\QATOPD. . {>}{<} }
%BeginExpansion
{> \atopwithdelims.. <}
%EndExpansion
0,\text{ } \\ 
\frac 1{\alpha -\omega \pm i\epsilon },\hspace{1.1in}\text{ }{\rm for}\text{ 
}\alpha \in \left( \omega _{\min },\omega _{\max }\right) , \\ 
\frac 1{z-\omega }\mp 2\pi i\delta (z-\omega ),\hspace{0.33in}\text{ }{\rm for%
}\text{ }{\rm Im}z%
%TCIMACRO{\QATOPD. . {<}{>} }
%BeginExpansion
{< \atopwithdelims.. >}
%EndExpansion
0.
\end{array}
\right.  \label{fz}
\end{equation}
$f_{\pm }(z)$ is analytic in all the complex plane. Let us check this
fact for $f_{+}(z).$ Approaching the real axis from above and below we have

\[
f_{+}(\alpha +i\epsilon )-f_{+}(\alpha -i\epsilon )=\frac 1{\alpha -\omega
+i\epsilon }-\frac 1{\alpha -\omega -i\epsilon }+2\pi i\delta (\alpha
-\omega ), 
\]
but that is exactly zero due to identity (\ref{ident}). Putting the
functions $f_{\pm }(z)$ into $R^{-1}(z)$ we get two analytic continuations, $%
R_{\pm }^{-1}(z)$ (see, e.g., Ref. \cite{Exner}). We can 
analytically continue the function $g^2(\alpha )$ across $\left( \omega _{\min
},\omega _{\max }\right) .$ We call $G(z)$ this extension, i.e. there is an
open region $\Delta$ of the complex plane containing $\left( \omega _{\min
},\omega _{\max }\right) $ and a meromorphic function $G:\Delta \rightarrow 
{\cal C}$ such that $g^2(\alpha )=G(\alpha )$ for $\alpha \in \left( \omega
_{\min },\omega _{\max }\right) .$ For notational convenience we write $%
g^2(z)=G(z)$ also for non-real $z.$ So, we are now ready to use the Cauchy
theorem and change the contour of integration in Eq. (\ref{ame}) by a
contour $\Sigma $ in the lower complex plane as shown in Fig. 27, which
leaves the singularities of $G(z)$ outside, namely

\begin{equation}
\left\langle B(t)\right\rangle =\int_\Sigma dz\frac{G(z)e^{-izt}}{%
R_{+}^{-1}(z)R_{-}^{-1}(z)}\left\langle B(0)\right\rangle .  \label{lcp}
\end{equation}

\epsfysize=10truecm
\centerline{\epsffile{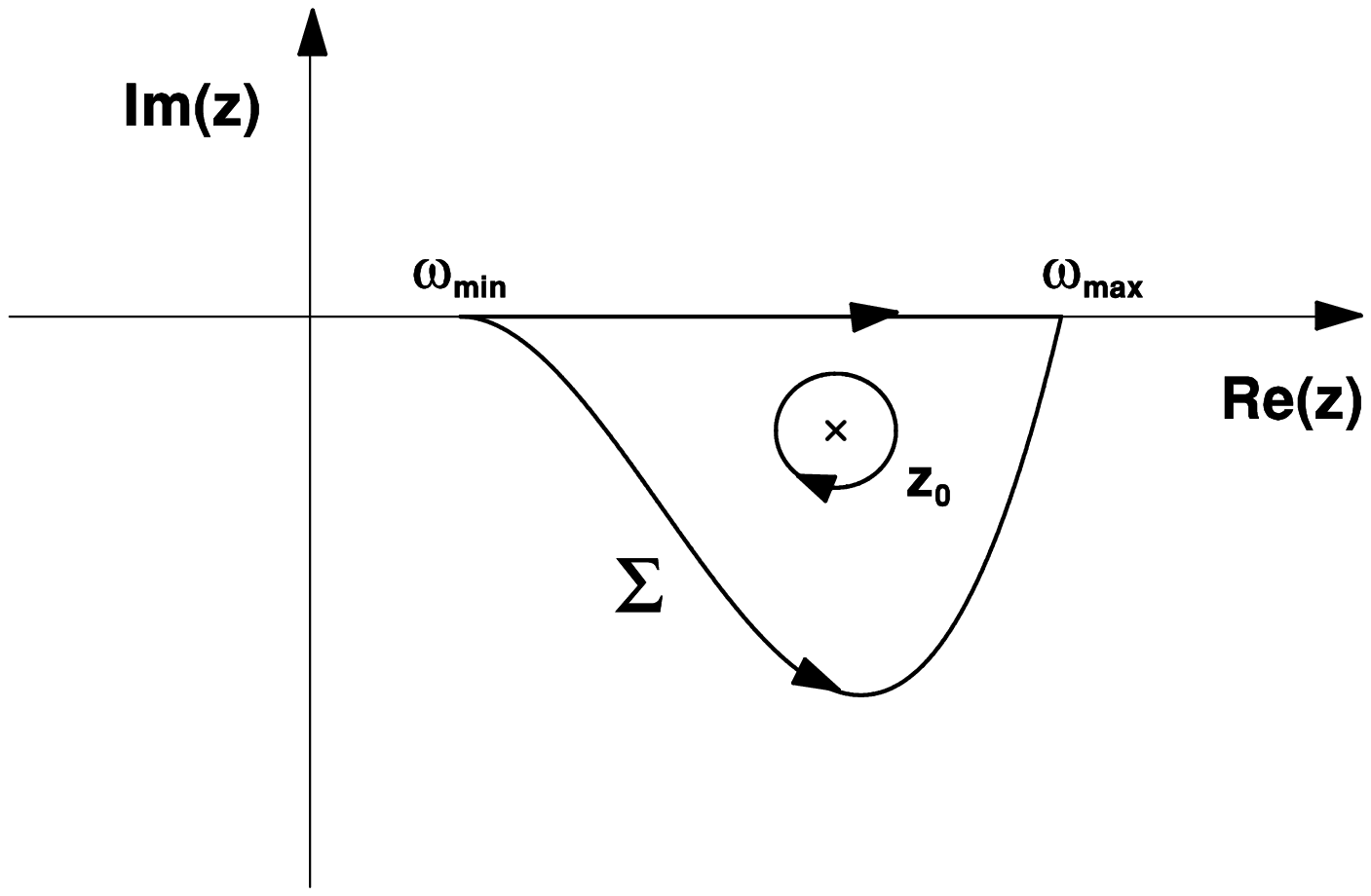}}
\centerline{{\bf FIG. 27.}\ Contour $\Sigma$.}
\bigskip
\bigskip

The last part of definition (\ref{fz}) can be considered as the values of $%
f(z)$ into the second sheet. The relevant fact is that $R_{+}^{-1}(z)$ has
now a complex zero formally given by the zero of 
\[
z-\Omega -\int_{\omega _{\min }}^{\omega _{\max }}d\omega \frac{g^2(\omega )%
}{z-\omega }+2i\pi G(z)=0. 
\]
This zero can be estimated up to the second order as

\[
z_0=\Omega +\int_{\omega _{\min }}^{\omega _{\max }}d\omega \frac{g^2(\omega
)}{\Omega -\omega -i\epsilon }+2i\pi G(\Omega )=\Omega +\delta \Omega -i%
\frac \Gamma 2. 
\]
That is, $\Omega $ has two corrections, a real one, which provides the
frequency shift, given by

\begin{equation}
\delta \Omega ={\rm PV}\int_{\omega _{\min }}^{\omega _{\max }}d\omega \frac{%
g^2(\omega )}{\Omega -\omega },  \label{shift}
\end{equation}
and a negative imaginary part, which represents the frequency width, given by

\begin{equation}
\Gamma =2\pi g^2(\Omega ).  \label{width}
\end{equation}
For the model we have developed in Sec. V this second order correction should 
vanish due to the symmetrical distribution of both the bath frequencies and
the interaction $g(\omega ).$

Coming back to our original purpose let us evaluate the integral of Eq. (\ref
{lcp}) for the case in which only a simple zero $z_0$ is present. By taking
the contour $\Sigma $ lying below $z_0$ and using the residues theorem we
have

\begin{equation}
\left\langle B(t)\right\rangle =\left[ \frac{e^{-iz_0t}}{\left(
R_{+}^{-1}\right) ^{\prime }(z_0)}+\int_{\omega _{\min }}^{\omega _{\max
}}d\alpha \frac{g^2(\alpha )}{\widetilde{R}_{+}^{-1}(\alpha
)R_{-}^{-1}(\alpha )}e^{-i\alpha t}\right] \left\langle B(0)\right\rangle ,
\label{as}
\end{equation}
where the tilde over $R_{+}^{-1}(z)$ stands for

\[
\frac 1{\widetilde{R}_{+}^{-1}(\alpha )}=\frac 1{R_{+}^{-1}(\alpha )}+2\pi i%
\frac{\delta (\alpha -z_0)}{\left( R_{+}^{-1}\right) ^{\prime }(z_0)}, 
\]
being $\delta (\alpha -z_0)$ the complex extension of the Dirac delta,
defined by

\[
\int_{\omega _{\min }}^{\omega _{\max }}d\alpha f(\alpha )\delta (\alpha
-z_0)=f(z_0),\ \ \ \ {\rm if}\ \ z_0\in {\rm Int}C_0, 
\]
which actually means $\frac 1{2\pi i}\oint_{C_0}dz^{\prime }\frac{%
f(z^{\prime })}{z^{\prime }-z_0}=f(z_0),$ for $C_0$ a contour encircling $%
z_0 $ into the second sheet as depicted by Fig. 28.

\epsfysize=10truecm
\centerline{\epsffile{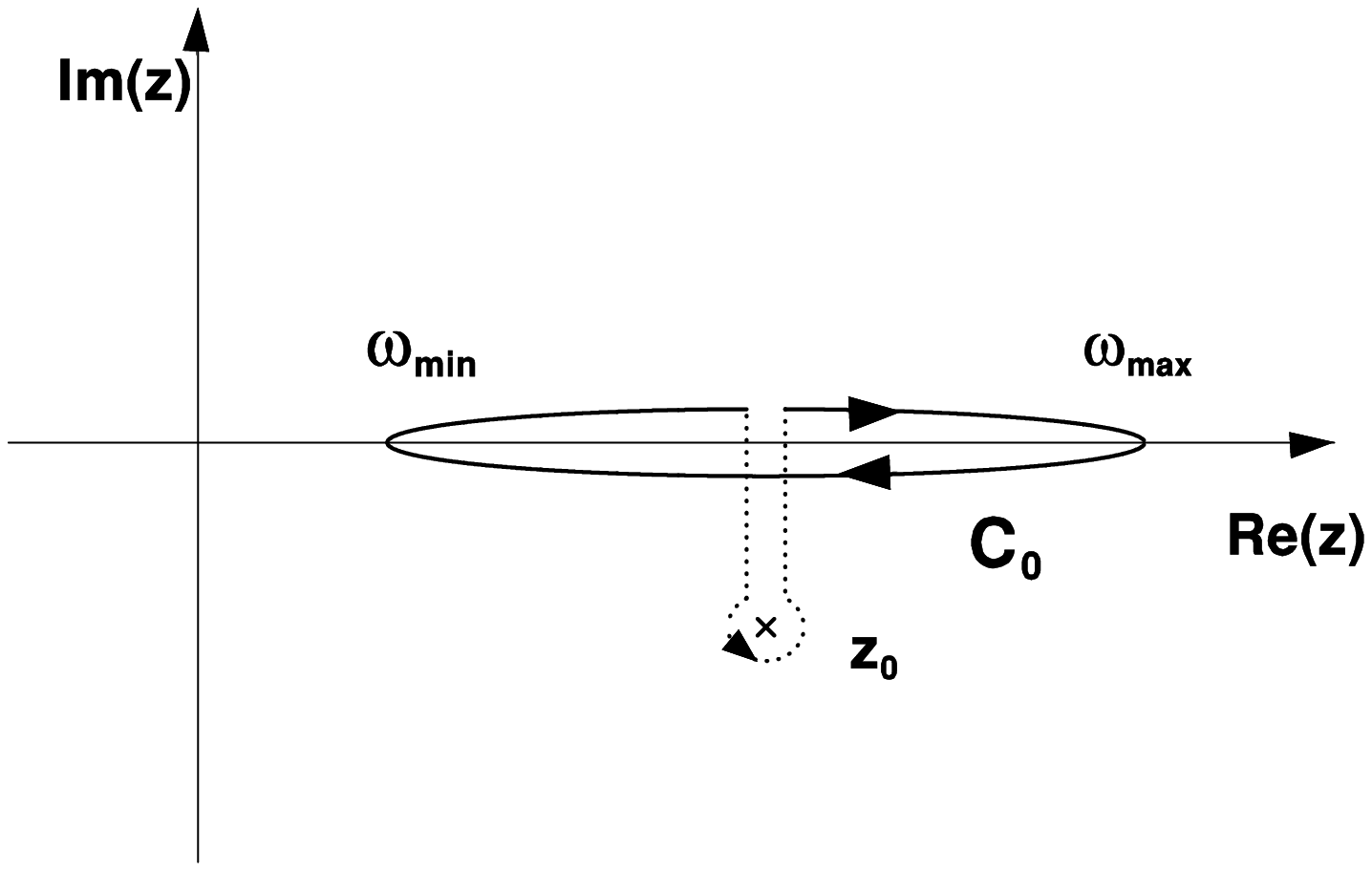}}
\centerline{{\bf FIG. 28.}\ Contour $C_0$; dotted line encircles $z_0$ into the 
second sheet.}
\bigskip
\bigskip

The second term inside
the brackets of Eq. (\ref{as}) is called the
background. It is responsible for deviations from the exponential law and
its contribution is relevant only for either very short or very long times.
In the regime where the exponential decay dominates we can neglect the
background. 
Then we obtain the approximate expression

\begin{equation}
\left\langle B(t)\right\rangle =\frac{e^{-iz_0t}}{\left( R_{+}^{-1}\right) ^{\prime }(z_0)}\left\langle B(0)\right\rangle ,
\label{i}
\end{equation}
which satisfies the following differential equation

\begin{equation}
\left\langle \stackrel{..}{B}(t)\right\rangle +z_0^2\left\langle
B(t)\right\rangle =0.  \label{zetacua}
\end{equation}
Eq. (\ref{zetacua}) has the form of a harmonic oscillator equation, however $%
z_0$ is now complex. Keeping in mind that $z_0=\Omega +\delta \Omega -i\frac 
\Gamma 2$, and neglecting the terms in $\Gamma ^2$, it is derived

\begin{equation}
\left\langle \stackrel{..}{B}(t)\right\rangle +(\Omega +\delta \Omega
)^2\left\langle B(t)\right\rangle +\Gamma \left\langle \stackrel{.}{B}%
(t)\right\rangle =0,  \label{Lang}
\end{equation}
which corresponds to the equation of a damped harmonic oscillator. Making
the same for the creation operator and since $X=1/\sqrt{2M\Omega }\left(
B+B^{\dagger }\right) $ we finally reach the standard form of the Langevin
equation (in mean values): 
\begin{equation}
\left\langle \stackrel{..}{X}(t)\right\rangle +(\Omega +\delta \Omega
)^2\left\langle X(t)\right\rangle +\Gamma \left\langle \stackrel{.}{X}%
(t)\right\rangle =0.  \label{chau}
\end{equation}
The first two terms on the left hand side represents the
Hamiltonian evolution of an oscillator with the renormalized frequency 
$\Omega +\delta \Omega,$
while the third one represents the `friction' (dissipative) part of the
damped linear oscillator with the damping factor $\Gamma.$ This $\Gamma$ is
the one we could have obtained if we had performed the same analytical continuation made in this section and if we had neglected the background in Eq. (\ref{coeclan}).
                      
\smallskip\ 

As we have pointed out in Eq. (\ref{surp}) the survival probability of the state $%
\left| \Omega \right\rangle $ is given by the square modulus of the integral
of Eq. (\ref{ame}). Let us study its short and long time behavior. From
general grounds, if we have a Hamiltonian given by $H=H_0+V,$ where $H_0$ is
an unperturbed Hamiltonian and $V$ a small perturbation, and $\left\{ \left|
\psi _n\right\rangle \right\} $ is a set of eigenvectors of $H_0,$ $%
H_0\left| \psi _n\right\rangle =E_n\left| \psi _n\right\rangle ,$ the
survival probability of the state $\left| \psi _k\right\rangle $ is given by 
$P_k(t)=\left| \left\langle \psi _k\right| \left. e^{-iHt}\psi
_k\right\rangle \right| ^2.$ At very short times the exponential can be
approximated by its first two terms in the Taylor expansion, $%
e^{-iHt}\approx 1-iHt.$ Then the survival probability behaves as $%
P_k(t)\approx 1+O(t^2),$ which does not correspond to an exponential decay
behavior like $1-\Gamma t.$ Thus, at very short times, we have a
non-exponential behavior, as it was shown is Sec. V, which is known as
Zeno's period \cite{Sudarshan}. For very long times we also have a
non-exponential contribution to the survival probability. As $t$ goes to
infinity the integral of Eq. (\ref{ame}) (survival amplitude) goes to zero
as a consequence of the Riemann-Lebesgue theorem. Then the behavior of the
survival probability depends on the small-frequency behavior of $g^2(\alpha
).$ For small frequencies $R^{-1}(\omega _{\min }+i\epsilon )=\omega _{\min
}-\Omega -\int_{\omega _{\min }}^{\omega _{\max }}d\omega \frac{g^2(\omega )%
}{\omega _{\min }-\omega +i\epsilon }\approx \omega _{\min }-\Omega ,$ since
the integral is bounded by condition (\ref{cpc}). The behavior of $g(\alpha
) $ is model dependent. We consider Ullersma's spectral strength of the kind 
$g(\alpha )=\frac{c_1\alpha }{\sqrt{c_2^2+\alpha ^2}},$ where $c_1$ and $c_2$
are constants, so the small-frequency behavior is given by $g(\alpha
)\approx (c_1/c_2)\alpha .$ Therefore we have, for $\frac 1{\omega _{\max }}%
\ll t<\frac 1{\omega _{\min }},$

\[
\int_{\omega _{\min }}^{\omega _{\max }}d\alpha \left| \Phi _\alpha \right|
^2e^{-i\alpha t}\approx \left( \frac{c_1}{c_2}\right) ^2\int_{\omega _{\min
}}^{1/t}d\alpha \left( \frac \alpha {\omega _{\min }-\Omega }\right)
^2e^{-i\alpha t}\sim \int_{\omega _{\min }}^{\omega _{\max }}d\alpha \alpha
^2e^{-i\alpha t}. 
\]
Calling $\alpha t=u$ we obtain for the survival amplitude

\[
\int_{\omega _{\min }}^{\omega _{\max }}d\alpha \alpha ^2e^{-i\alpha
t}=t^{-3}\int_{\omega _{\min }t}^{\omega _{\max
}t}duu^2e^{-iu}=it^{-3}\left. \left( u^2e^{-iu}-2iue^{-iu}-2e^{-iu}\right)
\right| _{\omega _{\min }t}^{\omega _{\max }t}. 
\]
Since the squared modulus behaves as $t^{-6}\left[ \left( \omega _{\max
}^4-\omega _{\min }^4\right) t^4+4\right] ,$ the survival probability gives
a power law decay

\begin{equation}
P_{\Omega \Omega }(t)\sim \Lambda ^4t^{-2}\hspace{0.3in}\text{ }{\rm as}%
\text{ }t\rightarrow \infty ,  \label{khalef}
\end{equation}
where $\Lambda =\sqrt[4]{\omega _{\max }^4-\omega _{\min }^4}$ is a measure
of an upper cutoff frequency. This deviation from the exponential decay law
given by a power series tail is known from Khalfin's original work \cite
{Khalfin}.

\smallskip\ 

Let us now study the asymptotic behavior of $\left\langle N_\Omega
(t)\right\rangle $ in the case of a dense bath, using the $\lambda ^2t$
approximation.
Going back to Eq. (\ref{fina}) and its complex conjugate we write

\begin{eqnarray*}
\left\langle N_\Omega (t)\right\rangle &=&\int_{\omega _{\min }}^{\omega
_{\max }}d\alpha \int_{\omega _{\min }}^{\omega _{\max }}d\alpha ^{\prime
}e^{i(\alpha -\alpha ^{\prime })t}\left| \Phi _\alpha \right| ^2\left| \Phi
_{\alpha ^{\prime }}\right| ^2\left\langle N_\Omega (0)\right\rangle \\
&&\ \ +\int_{\omega _{\min }}^{\omega _{\max }}d\alpha \int_{\omega _{\min
}}^{\omega _{\max }}d\alpha ^{\prime }e^{i(\alpha -\alpha ^{\prime })t}\psi
_\alpha \psi _{\alpha ^{\prime }}^{*}\int_{\omega _{\min }}^{\omega _{\max
}}d\omega \phi _\alpha ^{*}(\omega )\phi _{\alpha ^{\prime }}(\omega
)\left\langle N_\omega (0)\right\rangle ,
\end{eqnarray*}
where we have considered an uncorrelated initial state between subsystem and
bath, with the bath in thermal equilibrium, i.e. $\left\langle N_\omega
(0)\right\rangle =\left( e^{\beta \omega }-1\right) ^{-1}.$ From the
Riemann-Lebesgue theorem all the oscillating terms vanish as $t\rightarrow
\infty ,$ so the only survival term is that with the product of deltas
contained in $\phi ^{\prime }$s coefficients [Eq. (\ref{coefcl})], namely

\begin{equation}
\left\langle N_\Omega (\infty )\right\rangle =\int_{\omega _{\min }}^{\omega
_{\max }}d\alpha \frac{g^2(\alpha )}{\left| R_{+}^{-1}(\alpha )\right| ^2}%
\frac 1{e^{\beta \omega }-1}.  \label{asymp}
\end{equation}
We now perform the limit $\lambda \rightarrow 0.$ In such a limit $R_{\pm
}^{-1}(\alpha )\approx \alpha -\Omega \pm i\epsilon $ and taking into
account Eq. (\ref{sip}) we obtain

\begin{equation}
\left\langle N_\Omega (\infty )\right\rangle =\int_{\omega _{\min }}^{\omega
_{\max }}d\alpha \frac 1{2\pi i}\left[ \frac 1{\alpha -\Omega -i\epsilon }-%
\frac 1{\alpha -\Omega +i\epsilon }\right] \frac 1{e^{\beta \omega }-1}=%
\frac 1{e^{\beta \Omega }-1},  \label{asifin}
\end{equation}
since the expression between square brackets is $2\pi i\delta (\alpha
-\Omega ).$ We then see that the subsystem oscillator reaches thermal
equilibrium with the bath. In Sec. V we have studied the behavior of the
corresponding asymptotic form of $\left\langle N_\Omega \right\rangle $ for
a finite size bath [Eq. (\ref{asidis})]. In that case we have seen that,
when approaching to the continuous limit, $\sum_{\nu =0}^N\left( \frac{\Phi
_\nu ^2g_n}{\alpha _\nu -\omega _n}\right) ^2$ behaves as a delta
function. It is important to remark that Eq. (\ref{asifin}) shows that the
transfer of energy mainly occurs between two oscillators in resonance. In
Fig. 29 we have a diagrammatic representation of the time evolution of $%
\left\langle N_\Omega (t)\right\rangle .$ First, there is a deviation from
the exponential decay law due to the Zeno period, after that the exponential
decay dominates for long time the evolution until the Khalfin
power series tail, finishing in the asymptotic value of thermal
equilibrium given by Eq. (\ref{asifin}). Note that the continuous limit
carries Zeno time to zero and Khalfin time to infinity but nevertheless the
exponential decay law is not valid at all. Numerical estimates will be given
elsewhere.

\epsfysize=10truecm
\centerline{\epsffile{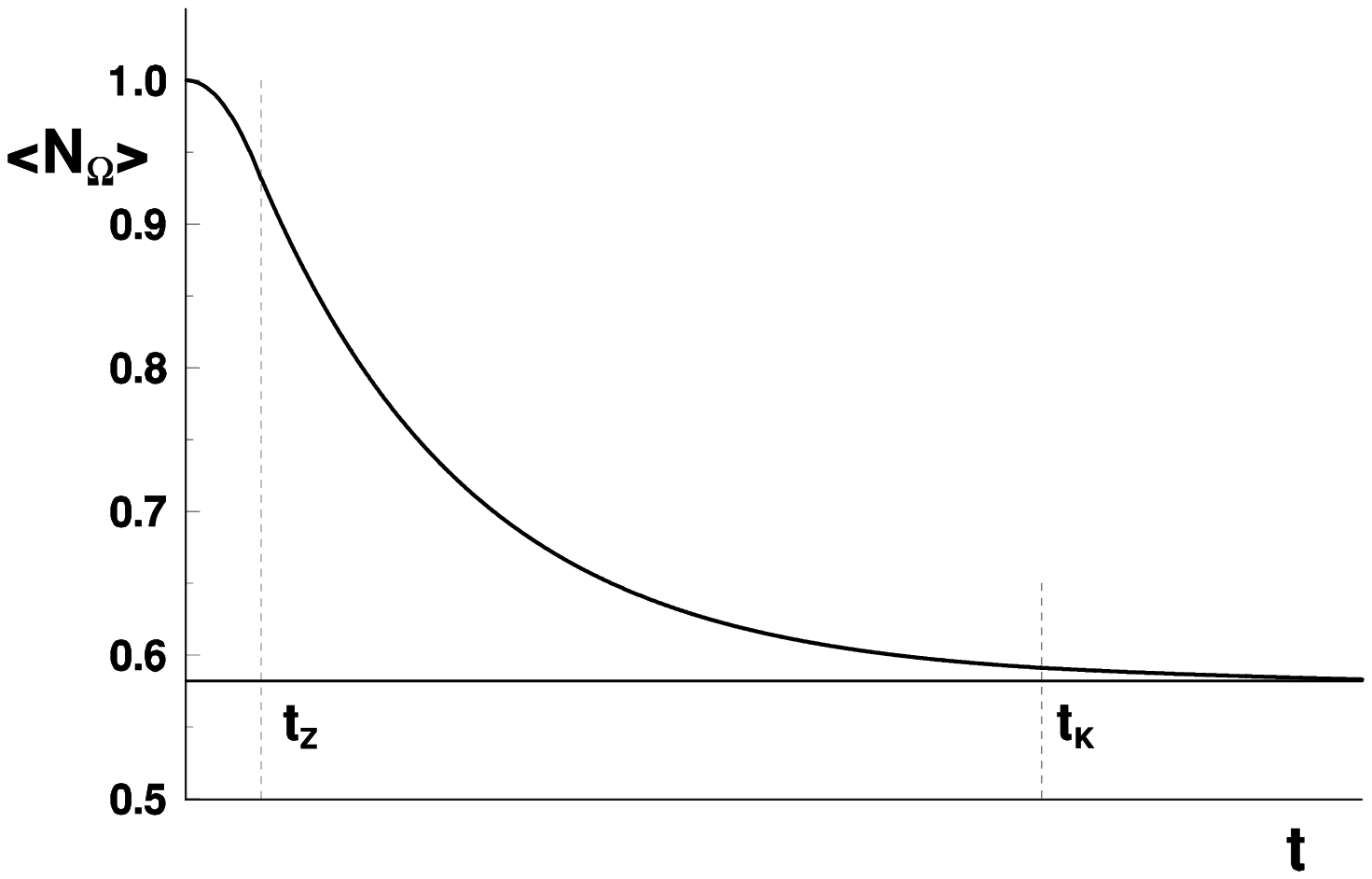}}
\centerline{{\bf FIG. 29.}\ Time scales.}
\bigskip
\bigskip

\medskip
\section{Concluding remarks}
\medskip

This work provides an exhaustive analysis of the most popular model of
Brownian motion in a way which has not been deeply explored in the literature on
the subject till now. The exact solution of the eigenvalue problem allows us to
study the time behavior of the magnitudes of interest without resorting to
approximations. No doubts about numerical errors can arise, since the
diagonalization method used, which has a powerful speed of calculus, does not
have recursive increasing deviations. Moreover the continuous limit is performed
in an analytic manner, obtaining the standard results found in the
literature. Figures of Sec. V clearly show all the properties expected for a
dissipative system (a damped oscillator in this case), with estimates of the
Poincar\'e recurrence time, fluctuations and equilibrium. The Poincar\'e
period arises as an exact time of revival and not as an statistical property
of the ensemble. The validity of the exponential decay law is also
enlightened. In a forthcoming paper we will study other statistical
properties of the model, such as correlation functions, and we will consider
other spectral densities and different ways to distribute the unperturbed
frequencies of the bath oscillators.

On the other hand, the problem of irreversibility can be traced in the
following way. Even for a finite system one can objectively `see' an
irreversible evolution, which only depends on the
system and not on the ability of the observer. This irreversibility is not a
consequence of a coarse-grained distribution or due to approximations.
Nevertheless  the time evolution is not strictly irreversible 
but it is practically
irreversible for our scale of observation. Moreover, if we consider a real
system with a large number of degrees of freedom (e.g. Avogadro's number),
it is easy to convince ourselves that we will see a time asymmetrical
evolution for the Brownian particle, since the Poincar\'e time becomes
larger than the age of the Universe. 

\acknowledgements

We are grateful to the organizers of the First International Colloquium on
`Actual Problems in Quantum Mechanics, Cosmology, and the Primordial 
Universe' and the `Foyer d'Humanisme' for their warm hospitality in Peyresq. 
We also thank Jos\'e Luis Gruver for useful discussions, 
and Ana Mar\'{\i}a Llois and Rub\'en Oscar Weht for reading the manuscript.


\begin{references}
\bibitem{Ullersma}  P. Ullersma, Physica {\bf 32}, 27-55 (1966).

\bibitem{Gordo}  M.A. Castagnino, F.H. Gaioli, and E. Gunzig, Fund. Cosmic
Phys. {\bf 16}, 221-375 (1996) and references therein.

\bibitem{Kozak}  R. Davidson and J.J. Kozak, J. Math. Phys. {\bf 12},
903-917 (1971).

\bibitem{Gruver}  J.L. Gruver, J. Aliaga, H.A. Cerdeira, and A.N. Proto,
Phys. Rev. E {\bf 51}, 6263-6266 (1995).

\bibitem{West}  K. Lindenberg and B.J. West, Phys. Rev. A {\bf 30}, 568-582
(1984) and references therein.

\bibitem{Friedrichs}  K.O. Friedrichs, Commun. Pure Appl. Math. {\bf 1},
361-406 (1948).

\bibitem{Haake}  F. Haake and R. Reibold, Phys. Rev. A {\bf 32}, 2462-2475
(1985).

\bibitem{van Hove}  L. van Hove, Physica {\bf 21}, 517-540 (1955). See also
E.B. Davies, Commun. Math. Phys. {\bf 33}, 171-186 (1973).

\bibitem{Sudarshan}  E.C.G. Sudarshan, C.B. Chiu, and V. Gorini, Phys. Rev.
D {\bf 18}, 2914-2929 (1978).

\bibitem{Laura} M.A. Castagnino and R. Laura, Phys. Rev. A {\bf 56}, 108-119 
(1997).

\bibitem{Exner}  P. Exner, {\it Open Quantum Systems and Feynman Integrals}
(Reidel, Amsterdam, 1985).

\bibitem{Khalfin}  L.A. Khalfin, Zh. Eksp. Teor. Fiz. {\bf 33}, 1371-1382
(1957) [Sov. Phys. JETP {\bf 6}, 1053-1063 (1958)].
\end{references}
\end{document}